\documentclass[journal=jpclcd,manuscript=letter]{achemso}

\usepackage{amsmath}
\usepackage{array}
\usepackage{graphicx}
\usepackage{amstext}
\usepackage{amsfonts}
\usepackage{amsmath,amssymb}
\usepackage{bm}
\usepackage{diagbox}

\newcommand{\be}{\begin{eqnarray}}
\newcommand{\ee}{\end{eqnarray}}
\newcommand{\ba}{\begin{array}}
\newcommand{\ea}{\end{array}}
\newcommand{\bml}{\begin{mathletters}}
\newcommand{\eml}{\end{mathletters}}

\newcommand{\lb}{\left(\begin{array}}
\newcommand{\rb}{\end{array}\right)}
\newcommand{\no}{\nonumber}

\newcommand{\dd}{\ddagger}

\date{\today}

\author{Pei-Yun Yang}
\affiliation{Department of Chemistry, Massachusetts Institute of Technology, Massachusetts, 02139 USA }
\author{Jianshu Cao}
\email{jianshu@mit.edu}
\affiliation{Department of Chemistry, Massachusetts Institute of Technology, Massachusetts, 02139 USA }

\title{Quantum Effects in Chemical Reactions under Polaritonic Vibrational Strong Coupling}

\begin{document}

\begin{tocentry}
\centerline{\scalebox{0.25}{\includegraphics[trim=0 0 100 80,clip]{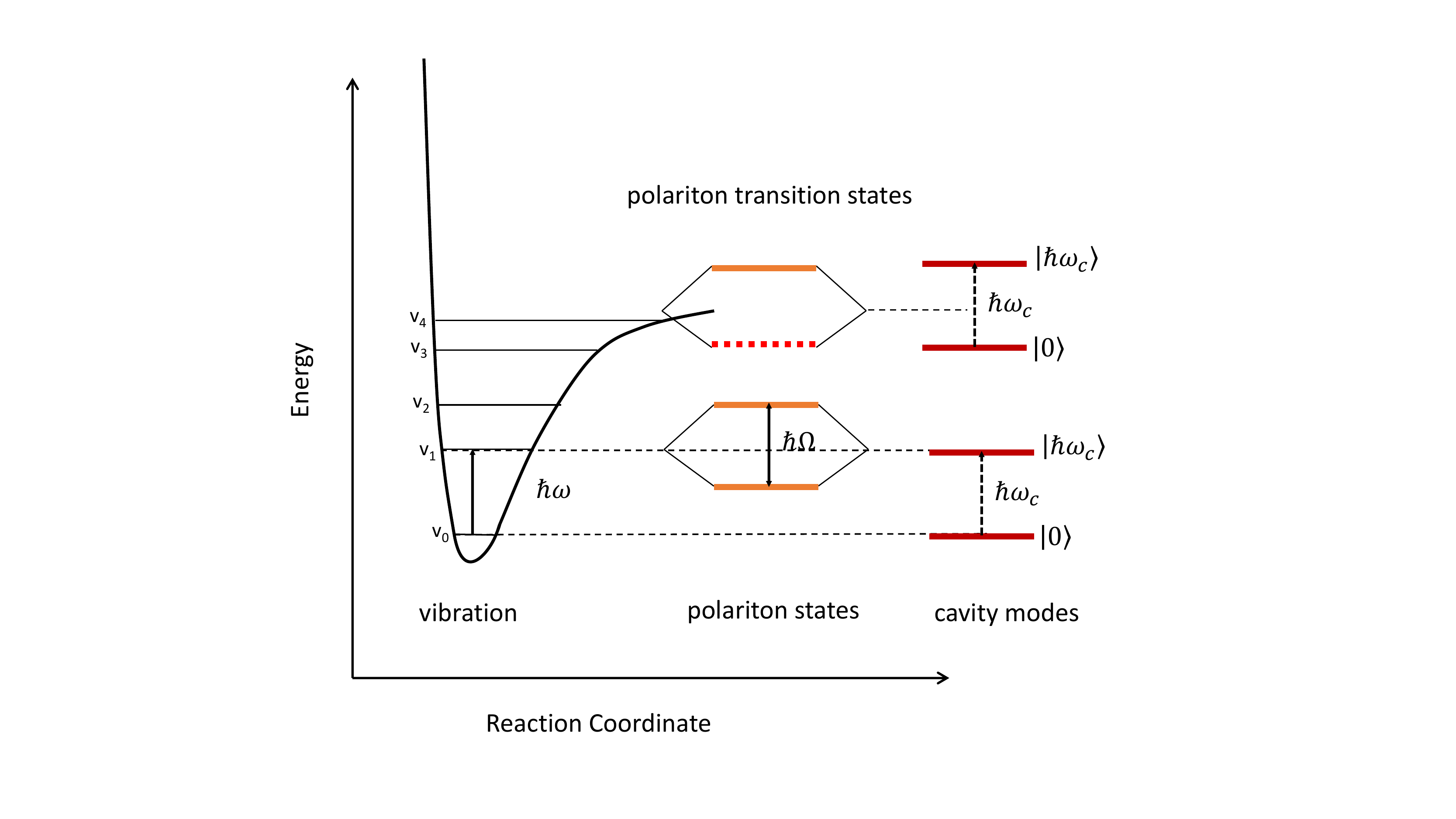}}}
\end{tocentry}

\begin{abstract}

The electromagnetic field in an optical cavity can dramatically modify and even control chemical reactivity via vibrational strong coupling (VSC). Since the typical vibration and cavity frequencies are considerably higher than thermal energy, it is essential to adopt a quantum description of cavity-catalyzed adiabatic chemical reactions. Using quantum transition state theory (TST), we examine the coherent nature of adiabatic reactions and derive the cavity-induced changes in eigen frequencies, zero-point-energy, and quantum tunneling. The resulting quantum TST calculation allows us to explain and predict the resonance effect (i.e., maximal kinetic modification via tuning the cavity frequency), collective effect (i.e., linear scaling with the molecular density), and selectivity (i.e., cavity-induced control of the branching ratio).  The TST calculation is further supported by perturbative analysis of polariton normal modes, which not only provides physical insights to cavity-catalyzed chemical reactions but also presents a general approach to treat other VSC phenomena.

\end{abstract}
\maketitle

\newpage

The electromagnetic field of an optical cavity can mix with quantum states of molecular systems to form polaritons and thus modify chemical kinetics in the vibrational strong coupling (VSC) regime.  In particular, recent experiments have clearly demonstrated dramatic effects on reactivity and selectivity of chemical kinetics when the vibrational and infrared (IR) cavity modes are strongly coupled.\cite{hutchison12,thomas16,ebbesen16,thomas19,lather19,lather21,hirai20a,hirai20b}
 Specifically, we highlight three effects:
 (i) {\bf resonance}, i.e., maximal cavity-induced suppression or enhancement of the reaction rate under the vibrational resonance condition; (ii) {\bf collectivity}, i.e. collective enhancement of the cavity-induced correction with the increase of molecule density in a cavity; (iii) {\bf selectivity}, i.e., regulation of the branching ratio of reaction channels via cavity frequency and other parameters.  These observations suggest the description of coherent light-matter interaction in terms of vibrational polaritons and support the physical picture where the IR mode catalyzes chemical reactions on the ground electronic state potential surface. This intriguing picture has stimulated theoretical and numerical studies;
yet the underlying mechanism of VSC-catalyzed chemical reactivity remains elusive.

Recent theoretical analysis adopts the quantum transition state theory (TST) for thermal rate calculations but has not fully elucidated the intriguing phenomena in VSC-catalyzed reactions.  With the TST framework, the timescale separation between electronic and nuclear degrees of freedom ensures that the reactions occur adiabatically on the ground-state potential energy surface, and the adiabatic reaction rate is predominantly determined by the activation energy on the ground state energy surface.\cite{li20a,angulo20}
Many intriguing molecular mechanisms such as dipolar interactions, anharmonicity, and anisotropic alignments 
have been proposed.\cite{galego19,climent19,triana20,vurgaftman20,schafer21}
In this paper, we examine the quantum nature of adiabatic reactions in cavities and analyze the resonant, collective, and selective effects of VSC-catalyzed reactions within the framework of multi-dimensional quantum TST.

{\textit{VSC Model and Partition Functions.}}
For simplicity, we begin with a single reactive molecule in a single mode cavity, where the vibrational strong coupling (VSC)
is described by the dipole interaction.\cite{flick17a,galego19}
Then, the Pauli-Fierz Hamiltonian \cite{power59} of  the cavity quantum electrodynamics (QED) is given as
\be
H =  H_{sys}+ H_{VSC} =  [{1 \over 2 } p^2 +  U(q)] + [ {1 \over 2} p_c^2 +  {1 \over 2}   (\omega_c q_c +  \mu(q) A_0 )^2]
\label{eq1}
\ee
where $\{p, q\}$ are the mass-scaled phase space variables of the reactive system,  i.e. $q= \sqrt{m}x$ and $p=\sqrt{m} \dot{x}$ with mass $m$,
 U(q) is the reactive potential,  $\{p_c, q_c\}$ are  the effective phase space variables of the cavity field,  $\omega_c$ is the cavity frequency, $\mu(q)$ is the dipole moment, and $A_0$ is the cavity potential strength.
The first term in the square brackets is the system Hamiltonian, $H_{sys}$,
and  the second term is the VSC Hamiltonian, $H_{VSC}$, which includes the dipole self-energy (DSE) term, $( \mu A_0)^2/2$.
Here, we assume a constant vector potential in the cavity such that,
$
A(q)= q A_0  = q / \sqrt{V \epsilon},  
$
where $V$  is the cavity volume and $\epsilon$ is the permitivity.
The QED potential A(q) is a vector and carries the polarization implicitly.  In this paper, we adopt a scalar notation,
 which can be easily translated to the vector format if needed.
 Further, all phase space variables and Hamiltonians are understood as quantum operators unless otherwise specified.

Interestingly, the form of light-matter interaction in Eq.~(\ref{eq1}) 
follows the same functional form as the Zwanzig Hamiltonian for a system embedded in a Gaussian bath,
where the classical thermodynamics of the open system is not affected by the bath. \cite{zwanzig73}
Explicitly, the reduced thermal distribution or, equivalently, the partition function of the  open system is given  in the classical limit  as
\be
Z^{cl}_{sys} =  {1\over Z^{cl}_{cav}} \int dq \int dp  \int dq_c \int dp_c  e^{-\beta H^{cl}}  =   \int dq \int dp   e^{-\beta H^{cl}_{sys}},
\label{classical}
\ee
where the superscript $cl$ denotes the classical limit and $Z^{cl}_{cav}$ is the classical partition function of an empty cavity.
Precisely because of the DSE term, the thermal distribution function of the system remains unperturbed by the cavity.  Thus, within this classical description,
all equilibrium properties, including bond length, dissociation energy, and activation energy, are not altered by VSC.\cite{li20a,angulo20,fischer21}
Yet, this argument becomes more subtle in the quantum version, also known as the Caldeira-Leggett Hamiltonian. 
In the  path integral formalism, the reduced quantum partition function becomes
\be
Z_{sys} = {1\over Z_{cav}} \int {\cal D} [q_c(\tau)] \int { \cal D} [q(\tau) ] e^{-\beta H}  =
 \int { \cal D} [q(\tau) ]  e^{-\int_0^\beta H_{sys} d\tau+ I[q(\tau)]}
\label{path}
\ee
which $I[q(\tau)]$ is  the influence functional.\cite{feynman00}
Physically, the influence functional arises from the quantized cavity field, which dresses the system under VSC,\cite{li20a}
and introduces corrections to equilibrium properties up to second order of light-matter coupling strength, i.e, $(\mu A_0)^2$, which is on the same order of the DSE term.\cite{fischer21}
These cavity-induced corrections can be evaluated perturbatively in the VSC regime, as illustrated in this paper. A direct effect of the influence functional is to introduce quantum fluctuations into the correlation function, which manifests in vibrational spectra.  These simple observations are consistent with a recent analysis of vibrational polaritons, in particular, regarding the role of the DSE term.\cite{fischer21}
For chemical reactions in a cavity, the influence functional modifies the zero-point-energy (ZPE) fluctuations and quantum tunneling effect,
which will be calculated in this paper within the framework of quantum transition state theory (TST).

To proceed, we expand the dipole moment around the equilibrium of the molecule potential (i.e. q=0, in the current notation), giving
$
\mu(q) = \mu_{eq} + \mu' q + \cdots    
$
where $\mu_{eq}$ is the permanent dipole at equilibrium
and $\mu' = (\partial \mu / \partial q)_{eq} $ is the gradient of the dipole, i.e., vibrational transition dipole moment.
 Due to the quadratic form of $H_{VSC}$,
the permanent dipole $\mu_{eq}$ term can be removed by shifting the cavity field according to
 $\omega_c q_c + \mu_{eq} A_0 \rightarrow \omega_c q_c$.
Thus, the cavity VSC Hamiltonian becomes independent of the permanent dipole and is rewritten as
\be
H_{VSC} =  {1 \over 2} p_c^2 +  {1 \over 2}  \omega_c^2 (q_c +  {\mu' A_0 \over \omega_c}  q  )^2
= {1 \over 2} p_c^2 +  {1 \over 2}  \omega_c^2 (q_c +  g q  )^2
\label{vsc}
\ee
where $g= {\mu' A_0 / \omega_c}$  is introduced as a dimensionless parameter to characterize the VSC strength.
Since experimental measurements are often reported in terms of the Rabi frequency $\Omega_R$, we establish the following relation,
$
\hbar \Omega_R =   g  \omega^2_c   \hbar /   \sqrt{ \omega \omega_c }   = 2 \hbar \omega_c \eta.
 $
 Here, $\eta$ is used to quantify the light-matter interaction strength.
In the Fabry-Perot cavity, the cavity frequency $\omega_c$ is tuned by changing the cavity length while keeping $\eta$ constant, so
we define the coupling constant $g$ as a function of  $\omega_c$,
\be
g= 2 \eta \sqrt{ \omega  \over \omega_c}
\ee
where $\eta$  and $\omega$ are fixed in the current setting.

\textit{ Quantum TST.}
A general starting point of quantum reaction rate is the stationary-phase approximation to the partition function,\cite{hanggi90,cao27} giving
\begin{align}
\label{stationary}
k_{TST} = {1\over \beta h} {Z_{\ddagger} \over Z_{eq}}
\end{align}
where $Z_{\ddagger}$ is the transition state (TS) partition function excluding the unstable mode associated with the reaction coordinate in the reactive barrier region and $Z_{eq}$ is the reactant partition function in the equilibrium well region.
We use $\ddagger$  to denote the unstable mode in the reactive barrier region (i.e. TS)  and associated quantities.
As explained in a comprehensive study,\cite{cao27} various rate expressions including transition state theory, centroid rate theory, and instanton
solution can be unified under the conceptual framework of Eq.~(\ref{stationary}).
 In this paper, we adopt the standard quantum transition state theory, as presented below.

We begin with the harmonic approximation for a single reaction coordinate and arrive at the quantum transition state theory (TST) expression
\begin{align}
k_{\rm TST} &=	\frac{2}{\beta h} \sinh\left(\frac{\omega\beta\hbar}{2}\right) e^{-\beta E_a}
\end{align}
where $\omega$ is the vibrational frequency in the reactant well and $E_a$ is the activation energy.  This is the case for
single molecules outside the cavity without the light-matter interaction.
Here, quantum tunneling is excluded and will be considered later in the context of centroid TST.
The single mode rate expression is generalized to the multi-dimensional TST rate $k^g_{TST}$  by introducing the correction factor $\kappa$,
defined via $k^{g}_{TST} = \kappa  k_{TST}$.   The correction factor can be written explicitly as\cite{wolynes81,pollak86a}
\begin{equation}
\kappa = { \prod_i \sinh(\lambda_i \beta \hbar /2)  \over \sinh(\omega \beta \hbar /2)  \prod^{\ddagger}_i \sinh(\lambda_{b,i} \beta \hbar /2) }
\label{kappa}
\end{equation}
where  $\lambda_i$ denotes the vibration eigen-frequency in the reactant well, $\lambda_{b,i}$ denote the vibrational eigen-frequency in the reactive barrier, 
and $\prod^{\dd}$ denotes all these modes excluding the unstable mode.   Specifically, in a cavity,  the coherent QED field couples
individual molecules, modifies the vibrational frequencies, and thus changes the rates,  which is quantified by the  the cavity-induced correction
factor $\kappa$.

To examine the cavity-induced effects on transition state rate, we rewrite the correction factor in Eq.~(\ref{kappa}) as
\begin{equation}
\kappa =	 { \prod_i  [1- \exp(-\lambda_i \beta \hbar ) ] \over   [1- \exp(-\omega \beta \hbar ) ]
\prod^{\ddagger}_i [1- \exp(\lambda_{b,i} \beta \hbar )]  } \exp(\beta \hbar S/2) = \kappa^* \exp(\beta \hbar S/2 ),
\label{kappa*}
\end{equation}
where $S$  is the frequency shift corresponding to the zero-point-energy (ZPE) contribution of vibrational modes to the activation energy,
given explicitly as,
\begin{equation}
S = \sum_i \lambda_i -\sum^{\ddagger}_i \lambda_{b,i} -\omega
\label{s}
\end{equation}
and $\kappa^*$ is the modified correction factor excluding the ZPE contribution.
In the high-temperature limit, the ZPE contribution can be ignored and the correction factor reduces to the classical limit,
\begin{align}
	 \kappa^*  \approx  \frac{1}{\omega} \frac{{\rm II}\lambda_i}{{\rm II}^{\ddagger}\lambda_{b,i}}
	=	\frac{\lambda_{\ddagger}}{\omega_{\ddagger}} = \kappa_{GH}
	\label{gh}
\end{align}
where $\omega_{\ddagger}$ is the unstable frequency without VSC, $\lambda_{\ddagger}$ is the unstable eigen-frequency under VSC,
and their ratio defines the Grote-Hynes (GH) factor  $\kappa_{GH}$.  The GH correction factor can be understood either as a multi-dimensional effect
on canonical TST or as a kinetic caging effect in generalized Langevin dynamics
 of the reaction coordinate.\cite{grote80,truhlar00,nitzan06,cao26}
 These two pictures are equivalent,
as demonstrated in Pollak's derivation of the Grote-Hynes rate based on the Zwanzig formalism
of dissipative dynamics.\cite{pollak86,pollak90}
In the low-temperature limit,  the multi-dimensional connection is dominated by the ZPE contribution and reduces to
\be
\kappa_{ZPE} \approx \exp( \beta \hbar S /2)
\label{zpe}
\ee
In between these two limits, a reasonable approximation is to combine the two limiting expressions, giving
$\kappa \approx \kappa_{GH} \exp(\beta \hbar S/2) $, which interpolates between the high and low temperature limits.
As shown in Fig.~1, the VSC-induced correction factor defined in Eq.~(\ref{kappa}) changes from the GH form in Eq.~(\ref{gh})
to the ZPE shift form in Eq.~(\ref{zpe}) as the temperature decreases.
Finally, to connect with thermodynamics, we can write Eq.~(\ref{kappa*}) as
\begin{align}
\kappa  \equiv
\underbrace{\kappa^*}_{entropy}exp{(\beta\underbrace{\hbar S/2) }_{enthalpy}}
\equiv \exp(-\beta \Delta  G_{\dd})
\label{thermal}
\end{align}
which allows us to identify $\kappa^*$ as the entropy correction
and the ZPE shift as the enthalpy correction and defines the cavity-induced free energy change $\Delta G_{\dd}$

\textit{ Single Molecule  Reaction in Cavity.}
In the framework of TST, the potential surface U(q) in Eq.~(\ref{eq1}) is approximated by a harmonic oscillator in the equilibrium reactant well, giving
\be
H =  H_s+ H_{VSC} =  [{1 \over 2 } p^2 +  {1 \over 2} \omega^2 q^2  ]+  [{1 \over 2} p_c^2 +  {1 \over 2}  \omega_c^2 (q_c + g q)^2]
\ee
where $\omega$ is the vibrational frequency at equilibrium.   The quadratic Hamiltonian defines the Hessian matrix,
\begin{align}
	\left[
	\begin{array}[c]{cc}
		\omega^2+g^2 \omega_c^2					& g \omega^2_c\\
		g \omega^2_c					&	\omega^2_c
	\end{array}\right]
	\notag
\end{align}
where $\omega^2 + g^2 \omega^2_c$ is the effective frequency of  the reaction coordinate.
Diagonalization of the Hessian matrix yields a pair of eigenvalues, $\lambda^2_+$ and $\lambda^2_-$,
\be
\lambda^2_{\pm} = {1\over 2} (\omega^2 +  g^2 \omega^2_c +\omega_c^2) \pm {1\over 2}
\sqrt{ (\omega^2+  g^2 \omega^2_c-\omega_c^2)^2 + 4 g^2 \omega_c^4}
\ee
In the perturbative regime, the two eigenfrequencies become
\be
\lambda_+ \approx  \omega_e + { (g\omega_c)^2 \omega \over 2  (\omega^2-\omega_c^2) }   \no \\
\lambda_- \approx  \omega_c - { (g\omega_c)^2 \omega_c \over 2 (\omega^2-\omega_c^2)} \no
\label{eigen}
\ee
which are nearly identical to the exact eigen-solution except at resonance $\omega=\omega_c$.
The divergence of the above perturbative expansion 
at the resonance is particularly interesting, as it suggests the largest perturbation due to VSC and thus the maximal
cavity-induced correction to the reaction rate [i.e. $\kappa$ in Eq.~(\ref{kappa})], which has been observed experimentally as the resonant effect.
Adding these two frequencies, we  have
\be
\lambda_+ +\lambda_-  =~\omega+\omega _{c}+\frac{g^{2}\omega _{c}^{2}}{2\left( \omega
+\omega _{c}\right) } +O\left( g^4 \right)
\ee
such that the overall frequency shift in the well is positive, indicating the increase of
the zero-point energy in the reactant due to VSC.
The detailed derivation and calibration of the perturbative solution can be found in Sec.~I of the supporting information (SI).

At the transition state, the potential energy surface can be approximated by a parabolic barrier, giving
$
U(q) = E_a  - {1\over 2}  \omega_{\ddagger}^2 q_{\dd}^2,
$
where $q_{\dd}$ is the mass-scaled barrier coordinate defined as $q_{\dd}=\sqrt{m} (x-x_{TS})$.
 The Hessian matrix at the barrier can now be written as
\begin{align}
	\left[
	\begin{array}[c]{cc}
		-\omega_{\ddagger}^2+  g_{\dd}^2 \omega^2_c				& g_{\dd} \omega^2_c\\
		g_{\dd} \omega^2_c					&	\omega^2_c
	\end{array} \right]
	\notag
\end{align}
where $\omega_{\ddagger}^2-g_{\dd}^2 \omega^2_c$ is the effective barrier frequency.
Again,  diagonalization of the above matrix yields a pair of eigenvalues,   $\lambda^2_{b+}$ and $-\lambda^2_{b-}$, which
take the same form as  Eq.~(\ref{eigen}), except for the replacement of $\omega^2$ with $-\omega^2_{\ddagger}$. i.e.,
\be
\pm \lambda^2_{b \pm} = {1\over 2} (\omega_c^2 - \omega_{\dd}^2 +  g_{\dd}^2 \omega^2_c) \pm {1\over 2}
\sqrt{ (\omega_{\ddagger}^2 -  g_{\dd}^2 \omega^2_c + \omega_c^2)^2 + 4 g_{\dd}^2 \omega_c^4}
\label{eigen-b}
\ee
The corresponding perturbative solutions are given as
\be
\lambda_{b} \equiv \lambda_{b+} \approx  \omega_c + { (g_{\dd}\omega_c)^2 \omega_c \over 2  (\omega_{\dd}^2 + \omega_c^2) } \no \\
\lambda_{\ddagger} \equiv \lambda_{b-}   \approx   \omega_{\dd} - { (g_{\dd}\omega_c)^2 \omega_{\dd} \over 2 (\omega_{\dd}^2 + \omega_c^2)} \no
\ee
which agree almost perfectly with the exact solution, as shown in Sec.~I of SI.
In the barrier region,  the stable frequency $\lambda_{b}$ increases, whereas the unstable frequency $\omega_{\ddagger}$ decreases;
both change quadratically with the cavity coupling strength $g_{\dd}$.

For a single reactive molecule in a cavity, the correction factor in Eq.~(\ref{kappa}) becomes
\begin{alignat}{2}
		\kappa 	=	\frac{\sinh\left(\frac{\lambda_+\beta}2\right)
														\sinh\left(\frac{\lambda_-\beta}2\right)}
												     {\sinh\left(\frac{\omega\beta}2\right)
														\sinh\left(\frac{\lambda_{b}\beta}2\right)}
\end{alignat}
The typical frequencies reported experimentally are higher than thermal energy, $\beta\hbar\omega > 1$, so the ZPE shift is the dominant contribution.
To leading order in $g$ and $g_\dd$, the ZPE shift is given as
\be
S = \lambda_+  + \lambda_- - \lambda_{b}-\omega
=  { \omega_c^3 \over 2} [ {g^2 \over \omega_c^2+\omega_c \omega} - {g_{\dd}^2 \over \omega_c^2+\omega_{\dd}^2}] + O(g^4)
\label{s-pert}
\ee
which is derived in Sec.~I of SI along with other perturbative results.
The perturbation expression determines the sign of the frequency shift and consequently the cavity-induced change in the reaction rate
\be
  S<0  \qquad e^{\beta \hbar S /2} <1 \qquad \text{if}  \qquad \omega\omega_c > \omega_{\dd}^2 \no \\
 S>0  \qquad e^{\beta \hbar S /2} >1 \qquad \text{if} \qquad  \omega \omega_c< \omega_{\dd}^2   \no
 \ee
 where $g=g_{\dd}$ is assumed.
 Evidently,  for a small cavity frequency, $\omega_c < \omega^2_{\dd} / \omega$,  the ZPE contribution suppresses the rate;
 otherwise, for a large cavity frequency,  $\omega_c < \omega^2_{\dd} / \omega$, the ZPE contribution enhances the rate.
 This is demonstrated in  the shift $S$ plotted in Fig.~1, which is positive at small $\omega_c$ and negative at large $\omega_c$.
 For the low cavity frequency,  $\kappa$ is not completely determined by the ZPE shift except at low temperature $\beta\omega\hbar=20$,
 so the rate is not necessarily suppressed even when  $\omega_c < \omega^2_{\dd} / \omega$, as shown in Fig.~1.
For typical reactive systems, we have $\omega_{\dd} < \omega$ and  $g_{\dd} > g$ such that the correction to the reaction rate is further suppressed in comparison with the special case of $g=g_\dd$ considered above.

\begin{figure}
\centerline{\scalebox{0.4}{\includegraphics[trim=0 0 0 0,clip]{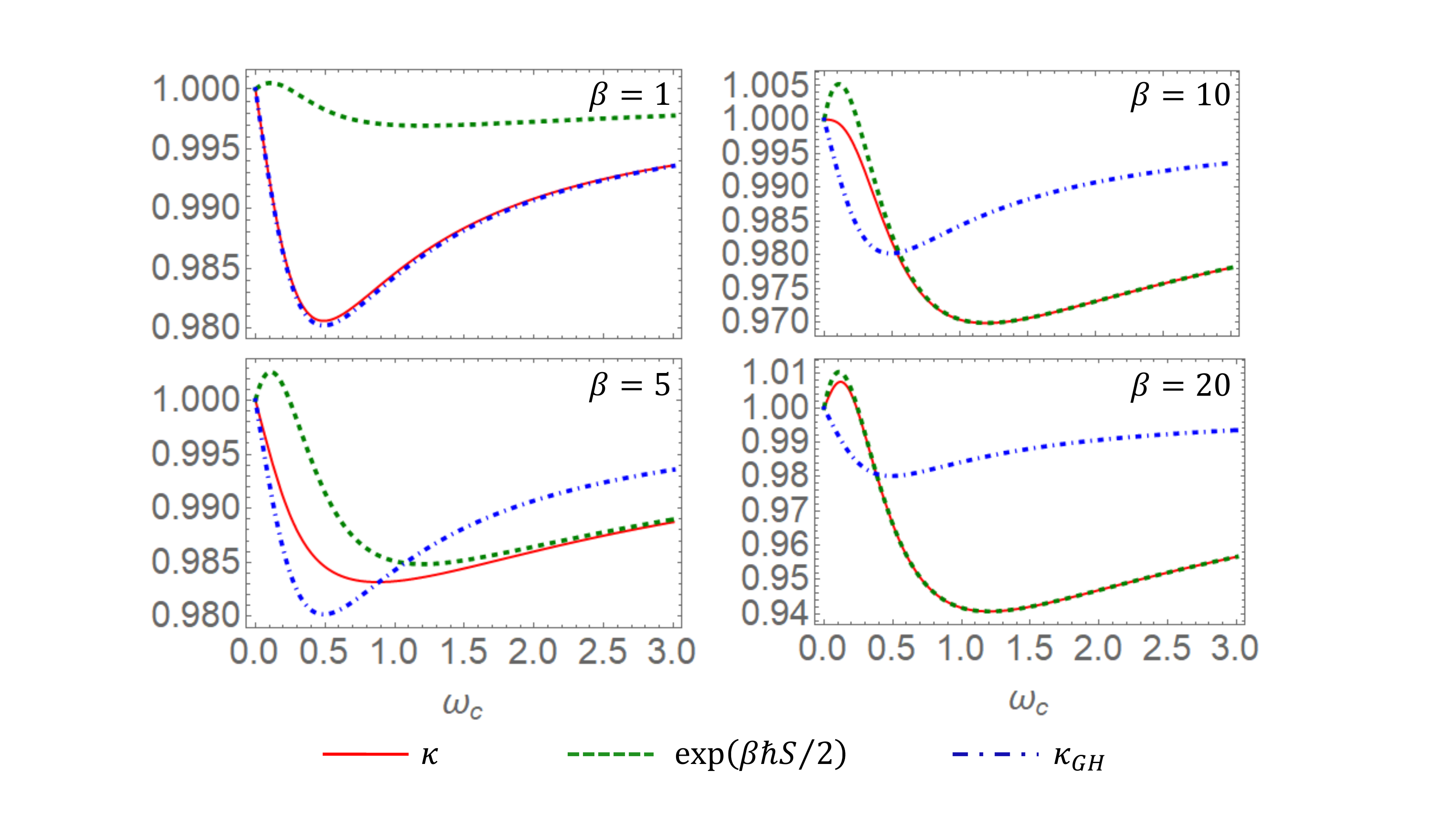}}}
\caption{The cavity-induced correction factor  $\kappa$  (solid lines) as a function of $\omega_c$ at different inverse temperatures $\beta$.
The high temperature limit $\kappa_{GH}$ (dot-dashed line) and the low-temperature limit $\kappa_{ZPE}$ (dashed line) are provided for comparison.
Relevant parameters are $\eta=\eta_\ddagger= 0.1$ and $\omega \ddagger=0.5$. All physical quantities are in unit of the vibrational frequency $\omega$.}
\label{Fig1}
\end{figure}

Fig.~1 compares VSC-induced correction factor $\kappa$ evaluated at different temperatures. We note the change of $\kappa$ from the high temperature limit
$\kappa\approx \kappa_{GH}$ to the low temperature limit $\kappa \approx \exp(-\beta \hbar S/2)$ as $\beta$ increases.  Accompanying this change,
Fig.~1  exhibits the shift of the resonance from the barrier frequency $\omega_c=\omega_\dd$ at high temperature to
near the vibrational resonance $\omega_c \approx \omega$ at the room temperature.
\begin{itemize}
\item
At high temperature,  i.e. $  \omega\hbar\beta \ll 1 $,
the reaction rate approaches the classical limit.  Then, according to Eq.~(\ref{kappa*}) and Eq.~(\ref{gh}),
 the cavity-induced correction reduces to the Grote-Hynes factor, i.e., $\kappa  \approx k_{GH}= \lambda_{\dd} / \omega_{\dd}$.
Using the leading order expression of Eq~(\ref{eigen-b}), we can easily identify the barrier resonant condition
$\omega_c=\omega_{\dd}$,\cite{li21a}  
where $\lambda_{\dd}$ or $\kappa_{GH}$ reaches the minimal as a function of cavity frequency (see Sec. I of SI).
In addition to the barrier resonance,   the classical correction factor  is always smaller than unity ($\kappa_{GH}<1$) and is temperature-independent.\cite{li21a}   
Evidently,  at room temperature, the typical vibrational energy gap is considerably larger than thermal energy; thus, classical TST is not applicable and a quantum mechanical treatment is essential. 
\item
At intermediate to low temperature, i.e., $ \omega\hbar \beta \ge 1 $, which is experimentally relevant, 
we use the exact expression in Eq.~(\ref{kappa}) and find  the resonance condition 
$\omega_c \approx  \omega$.   The vibrational resonance coincides with the curve crossing in Fig.~2(a), 
where the VSC has the maximal effect on the normal modes.
At low temperature,  $\kappa$ is dominated by the ZPE shift in Eq.~(\ref{zpe}), so the minimal in $\kappa$ can be approximately determined by  Eq.~(\ref{s-pert}).
In comparison,  the experiment measurement of the cavity-induced correction factor is larger, and the measured resonance width is narrower.
These discrepancies can arise from the collectivity and quantum tunneling effects, which are not considered in Figs.~1-3 but will be addressed 
in the later part of the paper.
 \end{itemize}

 \begin{figure}
\centerline{\scalebox{0.42}{\includegraphics[trim=0 0 0 0,clip]{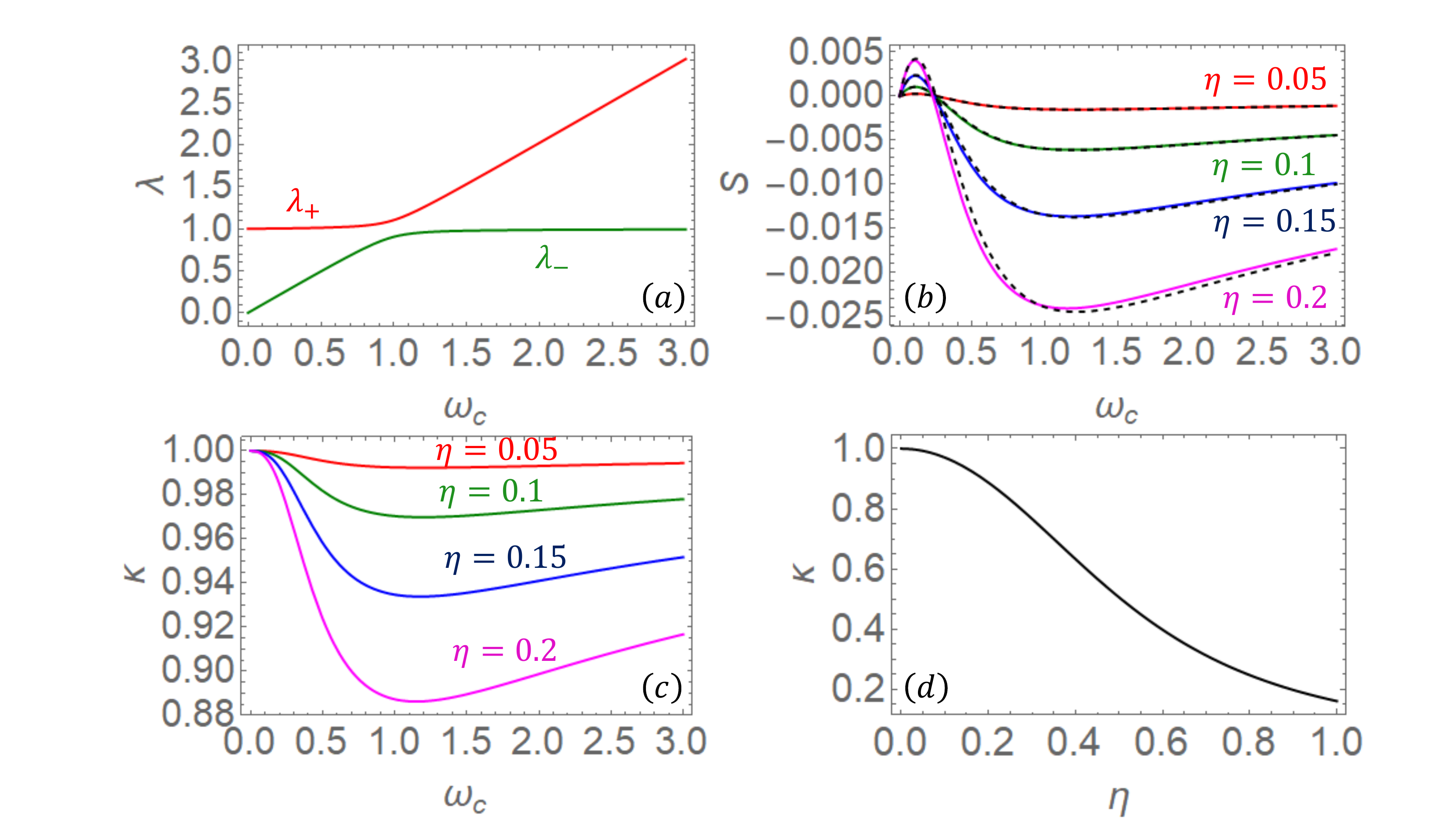}}}
\caption{(a) The two eigen frequencies in the reactant well, $\lambda_+$ and $\lambda_-$,  as a function of cavity frequency $\omega_c$
at the VSC strength of $\eta=0.1$;  (b) ZPE shift $S$ as a function of $\omega_c$ for different values of VSC strength $\eta$, 
along with the perturbation rsult (dashed lines);
(c) $\kappa$ as a function of $\omega_c$ for different values of VSC strength $\eta$;
(d)  $\kappa$ as a function of $\eta$ at  $\omega_c=1$.
All physical quantities are in unit of vibrational frequency $\omega$. 
Relevant parameters are  $\eta_{\ddagger}=\eta$, $\beta=10$, and $\omega_\ddagger=0.5$.}
\label{Fig2}
\end{figure}

Further,  as the VSC strength $\eta$ increases,  the ZPE shift in Fig.~2(b) and the correction factor $\kappa$ in Fig.~2(c) become amplified
while its resonance remains close to the vibrational frequency $\omega_c \approx \omega$.
In Fig.~2(b),  the perturbational expression of $S$ in Eq.~(\ref{s-pert}) is shown in good agreement with the exact solution.
Finally, In Fig.~2(d),  the $\eta$-dependence of $\kappa$
follows the perturbation analysis in Eq.~(\ref{s-pert}), which predicts $\ln (\kappa) \propto -\eta^2 \beta $.

\textit{ Mode Selectivity.}
Quantum control of reaction kinetics has been the holy grail of chemistry.
In a remarkable experiment,  Ebbesen and his coworkers\cite{thomas19}  have demonstrated
that the IR cavity can suppress or enhance a reaction channel relative to another channel and thus achieve chemical selectivity via tuning cavity frequency.
In other words, the VSC of a molecular system can lead to the preferential breaking or formation of one chemical bond among multiple IR active bonds through the VSC resonance.

Based on the normal mode analysis of single mode VSC in this paper,  we consider a reaction coordinate with two reactive barriers,
one to the left and another to the right of the reactant equilibrium.
Then, the branching ratio of the two reactive channels is
$
\phi_1 = { k_1 / ( k_1 + k_2) } = { 1 / (1+ k_2/k_1) }  
 $
which is determined by the ratio of rate constants.  Using Eq.~(\ref{kappa*}),  the ratio of the two TST rate constants is given as
\be
{k_2 \over k_1} =  { \kappa_2^*  \over  \kappa_1^*} \exp(\beta \hbar \Delta S/2 - \beta \Delta E_a)
\ee
where $\Delta S= S_2-S_1$ and $\Delta E_a= E_{a2}-E_{a1}$.
Since the reactant well is the same for both channels, the branching ratio depends on the nature of the barriers, characterized by three parameters
 $\{ \omega_{\dd},  g_{\dd}, E_a \} $.   Given the large  resonant frequency around $\omega \approx 10^3 \mbox{cm}^{-1}$,
 the cavity-induced correction factor is qualitatively described by the ZPE shift.  Then, using the perturbation result in Eq.~(\ref{s-pert}),
 the selectivity can be simply estimated by the ZPE difference,
 \be
\hbar {\Delta S \over 2} - \Delta E_a =
{\hbar \omega_c^3  \over 4}[ {g^2_{\dd 1}  \over \omega^2_c +\omega^2_{\dd 1} } - {g^2_{\dd 2}  \over \omega^2_c +\omega^2_{\dd 2}} ] +  (E_{a1}-E_{a2})
 \label{selectivity}
 \ee
 which is a function of cavity frequency $\omega_c$.

 For simplicity, we assume $E_{a1}=E_{a2}$, then the limiting values
  of selectivity  as shown in Fig.~3 are determined by two ratios: $\omega_{1\dd}/\omega_{2\dd}$ and $\eta_{1\dd}/\eta_{2\dd}$ .
 In the limit of small cavity frequency $\omega_c \rightarrow 0$, the sign of Eq.~(\ref{selectivity}) is determined by
 ${\eta_{\dd 1} / \omega_{\dd 1} }  - { \eta_{\dd 2} / \omega_{\dd 2} }$, which is negative for the parameters in Fig.~3,
 so the branching ratio favors channel 2, i.e., $\phi_1 < 50  \% $.
  For large cavity frequency,  the sign of   Eq.~(\ref{selectivity}) is determined by  $\eta_{1\dd} - \eta_{2\dd}$.
 Thus, in Fig.~3,   for $\eta_{2\dd}=0.12  > \eta_{1\dd} =0.1 $  and  $\eta_{2\dd}=0.11  > \eta_{1\dd} =0.1 $,
 the branching ratio  favors channel 1, i.e., $\phi_1 > 50  \% $
 at large cavity frequency and exhibits a switch of selectivity as a function of $\omega_c$.
 In comparison, the curve with  $\eta_{2\dd} = \eta_{1\dd} =0.1 $  always stay below $50 \%$  as the carrier frequency is tuned.
  In between the two limits, the branching ration exhibits a maximal reduction
 between the two barrier frequencies, $[ \omega_{1\dd}, \omega_{2\dd} ]$.
 
\begin{figure}
\centerline{\scalebox{0.5}{\includegraphics[trim=0 0 0 0,clip]{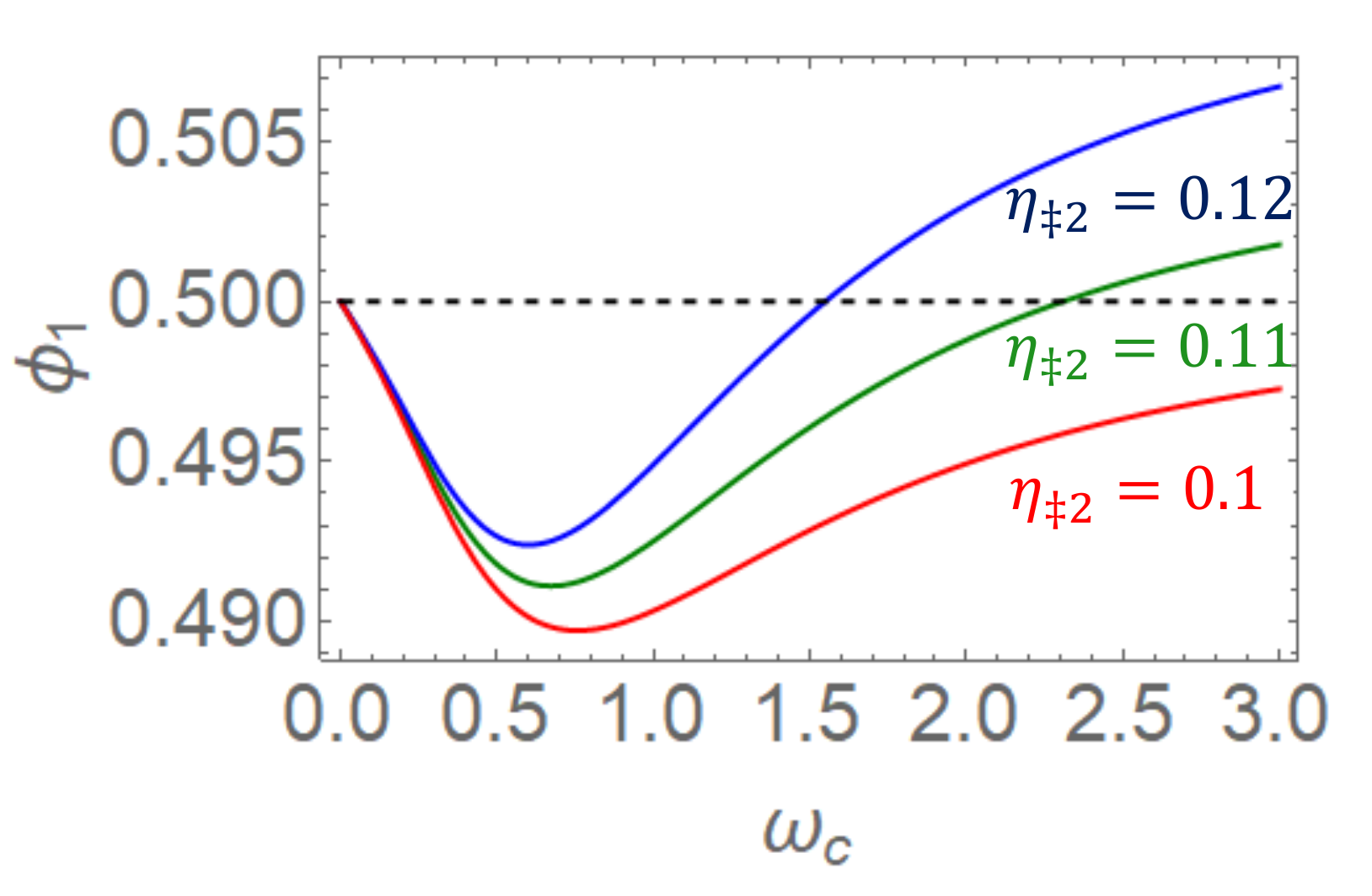}}}
\caption{Branching ratio  $\phi_1$ as a function of $\omega_c$ for three different values of  $\eta_{\ddagger 2}$.
Relevant parameters are  $\eta_1=\eta_{\ddagger 1} = 0.1$, $\omega_{\ddagger 1}=0.5$,
$\omega_{\ddagger 2}=1.2$, and $\beta=5$. All physical quantities are in unit of vibrational frequency $\omega$.}
\label{Fig3}
\end{figure}

Despite its simplicity, the single-mode model presented above confirms the possibility of cavity-enabled selectivity of chemical reactions.
To go beyond the single-mode picture,  one encounters
a more interesting problem of the non-linear coupling between molecular modes in a cavity, 
which bears similarity to the solvated ABA model.\cite{cao109}
 The correlated  mode couplings in the ABA model may shed light on the VSC-induced cooperativity
  in intramolecular vibrational relaxation (IVR), energy transfer, and reaction kinetics and will be a subject for future study.

\textit{ Collectivity and N-dependence.}
Experimental measurements demonstrate that both the Rabi frequency and cavity-induced correction increase with the molecular density.
This collective effect has inspired much theoretical interest\cite{gu20,angulo19}. Within the TST framework, individual molecules are
thermally activated without any coherence among molecules at the transition state such that the cavity correction $\kappa$
does not exhibit strong dependence on the molecular density.  
In contrast, in the coherent picture, the cavity creates coherent polariton states, which are thermally activated to react.  
Thus, the experimentally observed scaling with molecular density can be easily explained using
the N-scaled VSC coupling strength.  
Here, we first demonstrate the N-independence in the incoherent TST and then contrast it with the N-scaling in the coherent version.

To begin, we consider the N-particle Hamiltonian in a cavity,
\be
H =  H_{sys}+ H_{VSC} = [ \sum_{n=1}^N {1 \over 2 } p_n^2  +   U_N(q_1, \cdots, q_N) ]
 +  [{1 \over 2} p_c^2 +  {1 \over 2}  \omega_c^2 (q_c +  g \sum_{n=1}^N  q_n)^2]
 \label{N-hamiltonian}
\ee
where index $n$ is the particle index in the cavity and $U_N$ is the N-particle potential.
Here, we assume N non-interacting, identical molecules in the cavity.
In the reactant well,  we adopt the harmonic approximation,  $U_N = \sum_{n=1}^N  \omega^2 q_n^2/2 $,
so that the VSC ploariton is described by the Hessian matrix
\begin{align}
	\left[
	\begin{array}[c]{cc}
		\omega^2 + g^2_N \omega^2_c	&	g_{N} \omega^2_c\\
		g_N \omega^2_c					&	\omega^2_c
	\end{array}\right]
	\notag
\end{align}
where  $g_{N}=\sqrt{N} g$ is  the collective VSC constant for homogeneous coupling and can be easily generalized to the case of inhomogeneous coupling.
The Hessian matrix yields a pair of polariton frequencies, $\lambda_{\pm}$, and the remaining $N-1$ modes are dark states with
the unperturbed frequency $\omega$.

In the  incoherent  TST picture, one reactive molecule (e.g., $n=1$) is thermally activated to the transition state
while the other N-1 molecules remain in equilibrium,
so the TS potential can be approximated as
$
U_N =   E_a -  {1 \over 2}   \omega_{\dd}^2   q_{\dd}^2    + \sum_{n=2}^N   {1\over 2} \omega^2 q_n^2.
$
The corresponding Hessian matrix becomes
\begin{align}
	 \left[	
	\begin{array}{clr}
		-\omega^2_{\dd} + g_{\dd}^2 \omega^2_c	&	g_{N-1}  g_{\dd} \omega^2_c	& g_\dd \omega^2_c	\\
		g_{N-1} g_\dd \omega^2_c				&	\omega^2 + g^2_{N-1}  \omega^2_c	&	g_{N-1} \omega^2_c\\
		g_\dd \omega^2_c					&	g_{N-1} \omega^2_c			&\omega^2_c
	\end{array}\right]	
\notag
\end{align}
where the collective coupling is  $g_{N-1} = g \sqrt{N-1} $.
The eigensolution of the above Hessian matrix yields three eigenvalues: two real frequencies,
$\lambda^2_{b\pm}$, and one imaginary frequency, $-\lambda^2_{\dd}$.
The remaining $N-2$ modes are dark states with the equilibrium frequency $\omega$.
These eigenvalues are evaluated in Sec.~II of SI and yields the cavity-induced correction
\begin{alignat}{2}
	 \kappa  =	\frac{\sinh\left(\frac{\lambda_+\beta}2\right)
														\sinh\left(\frac{\lambda_-\beta}2\right)}
												{\sinh\left(\frac{\lambda_{b+}\beta}2\right)
														\sinh\left(\frac{\lambda_{b-}\beta}2\right)}
\end{alignat}
Surprisingly,  as shown in SI, to leading order of $g^2$,
 the energy shift $S(N)$ and barrier frequency $\lambda_\dd(N)$ of the N-particle system
 are identical to their corresponding values in the single molecule case, giving explicitly
 \be
 S(N)  & =&  ~ { \omega_c^3 \over 2} [ {g^2 \over \omega_c^2+\omega_c \omega} - {g_{\dd}^2 \over \omega_c^2+\omega_{\dd}^2}] + O(Ng^4)
		    \notag \\
\lambda_{\ddagger}(N) & =& ~ \omega_{\dd} [ 1- { (g_{\dd}\omega_c)^2  \over 2 (\omega_{\dd}^2 + \omega_c^2)} ]+ O(Ng^4)  \notag
 \ee
 where the first terms are exactly the simple molecule results and the next order corrections are in terms of $N g^4$.
Since there is no N-dependence in both the high-T and low-T limits of $\kappa$, we expect weak N-dependence in incoherent TST.
This prediction is confirmed in  Sec. II of SI and in Fig.~4(a), where $\kappa$ exhibits almost no N-dependence.

\begin{figure}
\centerline{\scalebox{0.5}{\includegraphics[trim=0 0 0 0,clip]{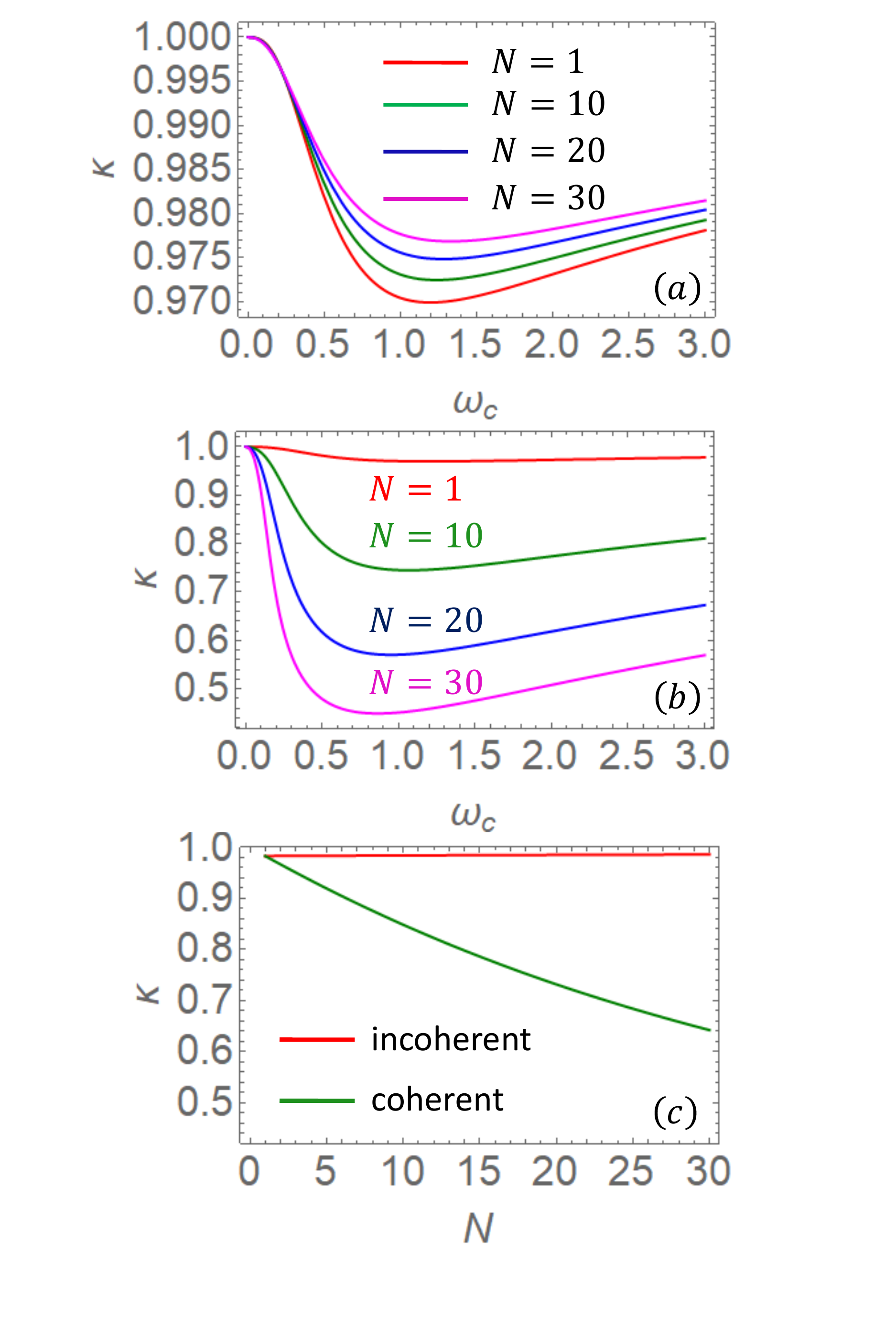}}}
\caption{(a)  $\kappa$ calculated in incoherent TST  as a function of $\omega_c$ for different $N$ with $\eta=\eta_\ddagger=0.1$; (b)
 $\kappa$ calculated in coherent TST  as a function of $\omega_c$  
 for different N with corresponding VSC strength $\eta=\eta_\ddagger=0.1\sqrt{N}$;
 (c) comparison of coherent and incoherent $\kappa$
  as a function of $N$ at $\omega_c=1$.  All  physical quantities are in unit of vibrational frequency $\omega$, and the other
 parameters are $\omega_\ddagger=0.5$ and $\beta=10$.}
\label{Fig4}
\end{figure}

In the coherent TST picture, the polariton is activated collectively to the transition state.  The resulting TS potential is 
 \be
U_N =   E_a -  {1 \over 2}   \omega_{\dd}^2   q_{\dd s}^2    + \sum_{i=2}^N   {1\over 2} \omega^2 q_i^2
\label{coherent}
\ee
where  $q_i$ are the i-th dark states and $q_{\dd s}$ is the bright state at the barrier, $q_{\dd s}=  \sum_n \sqrt{m_n} (x_n -x_{TS})/\sqrt{N} $. 
 Within the harmonic approximation, the above potential and the original molecular potential in Eq.~(\ref{N-hamiltonian}) are
related via a linear transformation and are thus equivalent.   The Hessian matrix at the TS is then given as
 \begin{align}
	\left[
	\begin{array}[c]{cc}
		-\omega_{\dd}^2 + g^2_{\dd N} \omega^2_c	&	g_{\dd N} \omega^2_c\\
		g_{\dd  N} \omega^2_c					&	\omega^2_c
	\end{array}\right]
	\notag
\end{align}
with $g_{\dd N}=g_\dd \sqrt{N}$. 
Therefore, all the results can be carried over from single-molecule reactions to N-particle coherent reactions if the VSC constant
g is scaled up by a factor of $\sqrt{N}$,  $g \rightarrow g\sqrt{N}$, which is exactly the same N-dependence for the Rabi frequency.   
To illustrate the coherent scaling,  Fig.~4(b) shows the dramatic enhancement of the cavity effect as N increases. 
Fig.~4(c) compares the coherent and incoherent
TST correction factors as a function of N and clearly shows the linear-scaling with N (equivalently, $\Omega_R^2$) in the coherent rate constant.

In coherent polariton theory, the collective VSC strength and cavity-induced correction can be enhanced as N increases. 
This type of cooperativity has been studied for non-adiabatic reactions in a cavity,\cite{galego17,angulo19,phuc20,gu20}
but not for adiabatic reactions on the ground-state surface.
In reality,  N should be interpreted as the number of coherently coupled molecules, which is limited by dynamic 
and static disorder.\cite{cao121,herrera16,scholes20}
Thus, the N-dependence in coherent reactions and other collective dynamics should be renormalized by 
localization and polaron effects.\cite{cao150,cao192}

\textit{ Quantum Tunneling.}
As explained in Eqs.~(\ref{kappa}) and (\ref{gh}),  the Grote-Hynes correction is a classical effect and the ZPE shift is a quantum effect.
Another contribution is quantum tunneling at the reactive barrier, which further enhances the cavity effect.
A simple way to account for the tunneling contribution is centroid TST,\cite{voth89}
which is deduced from the stationary phase evaluation of the barrier partition function in Eq.~(\ref{stationary})
using the centroid variable,\cite{cao27} giving
\be
k_{c-TST} = { (\omega_\dd \hbar\beta/2) \over \sin(\omega_\dd \hbar\beta/2) } k_{TST}  =  { \omega_\dd \over 2\pi }
{ \sinh(\omega\hbar\beta/2) \over \sin(\omega_\dd \hbar\beta/2) }  e^{-\beta E_a}
\ee
where $\omega_\dd$ and $\omega$ are the barrier and reactant frequencies, respectively.
The centroid TST rate is the same as quantum TST with the tunneling correction for a parabolic barrier.\cite{woynes81,pollak86a,cao27} 
With cavity VSC, the centroid rate becomes $k^{g}_{c-TST} = \kappa_{centroid} k_{c-TST}$, where the cavity correction factor in Eq.~(\ref{kappa}) becomes
\be
\kappa_{centroid} =  { \lambda_\dd \over  \omega_\dd}  { \sin(\omega_\dd \hbar\beta/2) \over \sin(\lambda_\dd \hbar\beta/2) }  \kappa
\label{kappac}
\ee
where $\kappa$ has been evaluated previously.  Fig.~5 shows $\kappa_{centroid}$ calculated from Eq.~(\ref{kappac}) at several temperatures.
In comparison with Eq.~(\ref{kappa}) and Fig.~1,
quantum tunneling dramatically enhances the cavity-induced correction and 
can be combined with the collective effect to improve the quantitative agreement with 
experimental measurements.

\begin{figure}
\centerline{\scalebox{0.5}{\includegraphics[trim=0 0 0 0,clip]{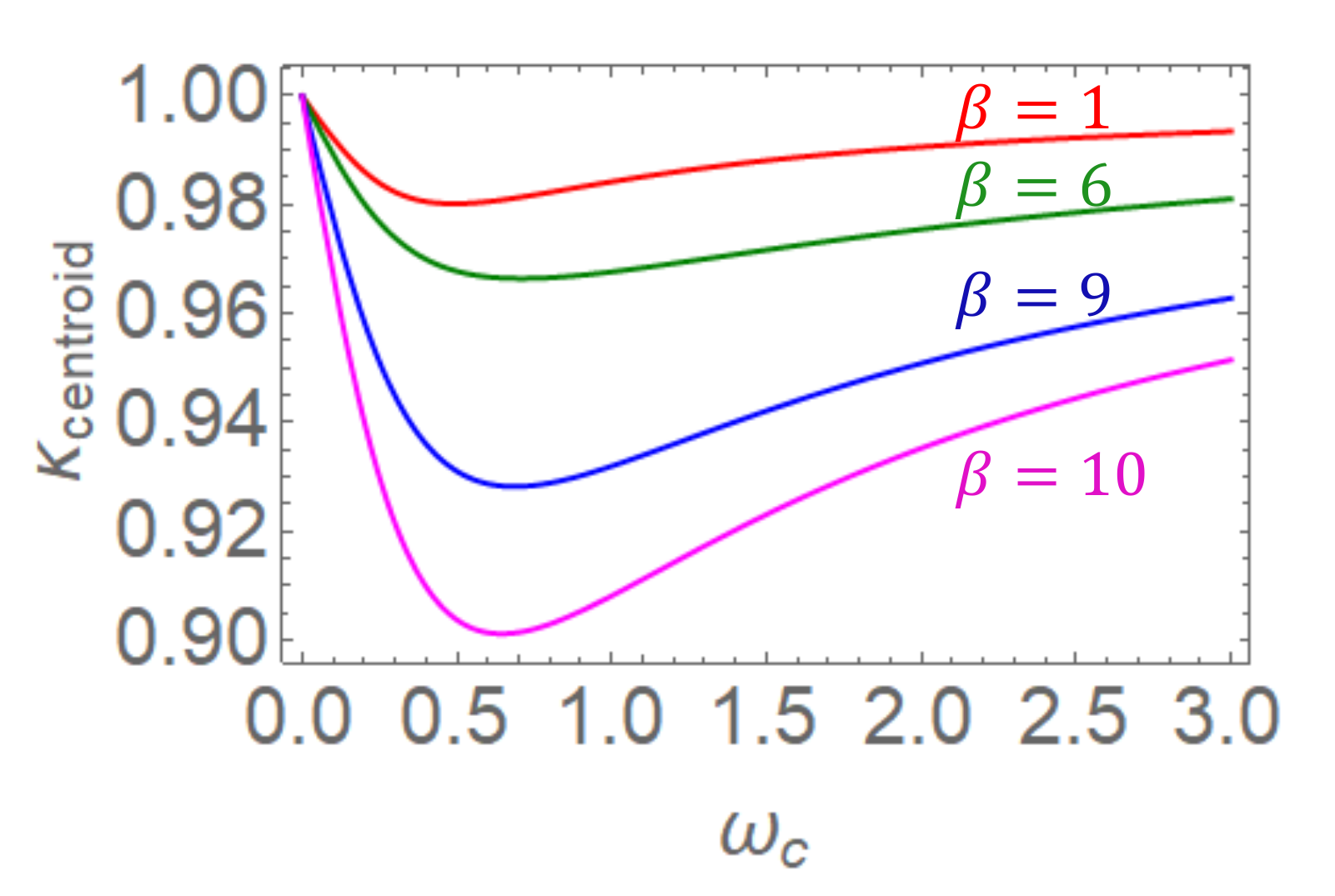}}}
\caption{The cavity-induced   correction factor  $\kappa_{centroid}$ defined in centroid TST as a function of $\omega_c$ at different temperatures.
All physical quantities are in unit of vibrational frequency $\omega$, and other relevant parameters are $\eta=\eta_\dagger=0.1$
and $\omega_\ddagger=0.5$. }
\label{Fig5}
\end{figure}

Interestingly, the pre-factor of the centroid TST diverges as $\omega_\dd \hbar\beta_c= 2\pi$, which suggests the failure of TST 
at or below the crossover temerature $T_c$.
An easy fix is to replace the anharmonic well and barrier potentials with the optimal quadratic free energy potentials determined
by the variational principle.\cite{cao26,cao19}
More generally,  in the deep tunneling regime (i.e. $T<T_c$),  quantum fluctuations dominate and the concept of 'transition state'
based on a local reaction coordinate is no longer valid.  Instead, we should adopt the picture of 'coherent tunneling', which is delocalized
and involves the full anharmonic potential.  One approach is based on the concept of 'instanton',
which is a nontrivial stationary solution to the barrier partition function at $T<T_c$ and predicts the quantum tunneling rate.\cite{miller75,cao27}
 As stated earlier, the cavity field can be treated as a single-mode harmonic bath
and thus introduces an influence functional in the path integral expression of the system partition function [see Eq.~(\ref{path})].
 In a previous study,
 the harmonic bath effect on the tunneling rate was found to reduce the instanton rate as the system-cavity coupling increases,\cite{cao27,cao21}  consistent with the experimental measurements in optical cavities. Further, quantum tunneling can permit a natural justification of the coherence and collectivity as described in Eq.~(\ref{coherent}).  

In summary, 
as the typical vibration and cavity frequencies $(10^3 cm^{-1})$
are considerably higher than thermal energy, it is essential to adopt a quantum description of ground-state
 chemical reactions in a cavity. Within the framework of quantum TST,  we are able to account for zero-point energy shift and quantum tunneling
 induced by VSC and demonstrate the resonant effect, collective effect, and selectivity,  as well as various parametric dependences as summarized below. 
 Based on the relatively simple TST calculations, our predictions are consistent with experiments.  Much of the reported study is built on perturbative normal mode analysis, which not only provides physical insights into cavity-catalyzed chemical reactions but also presents a useful tool to treat other VSC effects.
\begin{enumerate}
\item  {\it Dipole self-energy term and thermal equilibrium:}
The DSE term introduces additional restoring forces along the reaction coordinate and effectively increases the vibrational frequency.  According to the perturbation analysis,  the DSE contribution is on the same order as the linear dipole term in the PF Hamiltonian and thus cannot be ignored in the VSC regime.  As a result of the Zwanzig form of the VSC hamiltonian, the equilibrium thermodynamics of the cavity-dressed system exhibits no VSC perturbation 
in the classical regime  (see Eq.~(\ref{classical}))  and a temporal correlation in the quantum regime  (see Eq.~(\ref{path})).  
This explains the weak cavity effects on equilibrium quantities including the bond length, disassociation energy, and activation energy.  For the same reason, the permanent dipole does not contribute to the thermal reaction rate and the leading order is the linear transition dipole moment (see Eq.~(\ref{vsc})). 
\item {\it Cavity-catalyzed single molecule reactions:}
(i) The cavity-induced correction can be attributed to the ZPE shift quantified by $S$ and the kinetic effect quantified by $\kappa^*$, which can be related to enthalpy and entropy changes respectively (i.e. Eq.~(\ref{thermal})).
In the experimentally relevant regime, the dominant contribution arises from the ZPE shift except for small cavity frequency.
(ii) The sign of ZPE shift in Eq.~(\ref{s-pert}) predicts the suppression or enhancement of the reaction rate, and the amplitude of the shift depends
  on the VSC strength or the Rabi frequency quadratically. (iii) With typical parameters, the cavity-induced correction exhibits a maximal supression
   when the cavity frequency is near the vibrational resonance frequency.
\item {\it Mode selectivity:}
VSC can either enhance or suppress the branching ratio in a multiple-channel reaction, thus achieving mode selectivity in a cavity.
Within a single-mode two-barrier model, the branching ratio of the two competing reactions
is determined by the relative free energy difference of the two reactive barriers, i.e. Eq.~(\ref{selectivity}),
characterized by three parameters $\{ \omega_{\dd},  g_{\dd}, E_a \} $.   Perturbation analysis establishes the functional-dependence of the branching
ratio on these parameters and suggests the reordering of preferential selectivity as the cavity frequency $\omega_c$ is tuned.
\item{\it Collectivity and N-Scaling:}
In the TST framework,  
individual molecules are thermally activated without any coherence between molecules at the transition state. Within this incoherent picture,  to leading order of VSC strength,  the frequency shift and the correction factor are identical to the single molecule results and are thus N-independent.
 In contrast, in the coherent picture,  the polariton state is thermally activated to yield collective barrier crossing. 
 Thus, the experimentally observed collectivity can be easily explained by rescaling the single molecule VSC strength, $g(N)=\sqrt{N}g$. 
\item{\it Quantum tunneling:}
Centroid TST is used to account for quantum tunneling at the reactive barrier for $T>T_c$ 
and is shown to enhance the VSC-induced correction considerably.
Yet, the centroid correction diverges at the crossover temperature $T_c$, which indicates the failure of TST at $T \le T_c$ 
 and the emergence of coherent tunneling as described in the instanton picture, where under-barrier tunneling dominates over
 thermal activation.
 \end{enumerate}
Chemical reactions are complex, involving cohesive motion of many degrees of freedom in a thermal environment.\cite{garcia21}
Theoretical models based on normal mode analysis allow for analytical solutions and provide useful insight into this complex process.
Part of the simplicity arises from the
harmonic form of the cavity field, which introduces a tunable bosonic bath mode.  Therefore,
many emerging phenomena in polariton chemistry\cite{garcia21} can be understood with the 
combination of  quantum simulations and analytical solutions.

\section*{Acknowledgement} This work is supported by NSF (CHE 1800301 and CHE 1836913).

\section*{Supplementary Information}
The Supporting Information is available: S1. Perturbation Analysis: Single Molecule VSC;
S2.  Perturbation Analysis: N-Molecule VSC

\newpage

\bibliography{./bibfiles/polariton,./bibfiles/pub_cao}

\providecommand{\latin}[1]{#1}
\makeatletter
\providecommand{\doi}
  {\begingroup\let\do\@makeother\dospecials
  \catcode`\{=1 \catcode`\}=2 \doi@aux}
\providecommand{\doi@aux}[1]{\endgroup\texttt{#1}}
\makeatother
\providecommand*\mcitethebibliography{\thebibliography}
\csname @ifundefined\endcsname{endmcitethebibliography}
  {\let\endmcitethebibliography\endthebibliography}{}
\begin{mcitethebibliography}{47}
\providecommand*\natexlab[1]{#1}
\providecommand*\mciteSetBstSublistMode[1]{}
\providecommand*\mciteSetBstMaxWidthForm[2]{}
\providecommand*\mciteBstWouldAddEndPuncttrue
  {\def\EndOfBibitem{\unskip.}}
\providecommand*\mciteBstWouldAddEndPunctfalse
  {\let\EndOfBibitem\relax}
\providecommand*\mciteSetBstMidEndSepPunct[3]{}
\providecommand*\mciteSetBstSublistLabelBeginEnd[3]{}
\providecommand*\EndOfBibitem{}
\mciteSetBstSublistMode{f}
\mciteSetBstMaxWidthForm{subitem}{(\alph{mcitesubitemcount})}
\mciteSetBstSublistLabelBeginEnd
  {\mcitemaxwidthsubitemform\space}
  {\relax}
  {\relax}

\bibitem[Hutchison \latin{et~al.}(2012)Hutchison, Schwartz, Genet, Devaux, and
  Ebbesen]{hutchison12}
Hutchison,~J.~A.; Schwartz,~T.; Genet,~C.; Devaux,~E.; Ebbesen,~T.~W. Modifying
  chemical landscapes by coupling to vacuum fields. \emph{Angew. Chem. Int.
  Ed.} \textbf{2012}, \emph{51}, 1592\relax
\mciteBstWouldAddEndPuncttrue
\mciteSetBstMidEndSepPunct{\mcitedefaultmidpunct}
{\mcitedefaultendpunct}{\mcitedefaultseppunct}\relax
\EndOfBibitem
\bibitem[Thomas \latin{et~al.}(2016)Thomas, George, Shalabney, Dryzhakov,
  Varma, Moran, Chervy, Zhong, Devaux, Genet, Hutchison, and Ebbesen]{thomas16}
Thomas,~A.; George,~J.; Shalabney,~A.; Dryzhakov,~M.; Varma,~S.~J.; Moran,~J.;
  Chervy,~T.; Zhong,~X.; Devaux,~E.; Genet,~C. \latin{et~al.}  Ground-state
  chemical reactivity under vibrational coupling to the vacuum electromagnetic
  field. \emph{Angew. Chem.} \textbf{2016}, \emph{128}, 11634--11638\relax
\mciteBstWouldAddEndPuncttrue
\mciteSetBstMidEndSepPunct{\mcitedefaultmidpunct}
{\mcitedefaultendpunct}{\mcitedefaultseppunct}\relax
\EndOfBibitem
\bibitem[Ebbesen(2016)]{ebbesen16}
Ebbesen,~T.~W. Hybrid light-matter states in a molecular and material science
  perspective. \emph{Acc. Chem. Res.} \textbf{2016}, \emph{49}, 2403\relax
\mciteBstWouldAddEndPuncttrue
\mciteSetBstMidEndSepPunct{\mcitedefaultmidpunct}
{\mcitedefaultendpunct}{\mcitedefaultseppunct}\relax
\EndOfBibitem
\bibitem[Thomas \latin{et~al.}(2019)Thomas, Lethuillier-Karl, Nagarajan,
  Vergauwe, George, Chervy, Shalabney, Devaux, Genet, Moran, and
  Ebbesen]{thomas19}
Thomas,~A.; Lethuillier-Karl,~L.; Nagarajan,~K.; Vergauwe,~R.~M.; George,~J.;
  Chervy,~T.; Shalabney,~A.; Devaux,~E.; Genet,~C.; Moran,~J. \latin{et~al.}
  Tilting a ground-state reactivity landscape by vibrational strong coupling.
  \emph{Science} \textbf{2019}, \emph{363}, 615\relax
\mciteBstWouldAddEndPuncttrue
\mciteSetBstMidEndSepPunct{\mcitedefaultmidpunct}
{\mcitedefaultendpunct}{\mcitedefaultseppunct}\relax
\EndOfBibitem
\bibitem[Lather \latin{et~al.}(2019)Lather, Bhatt, Thomas, Ebbesen, and
  George]{lather19}
Lather,~J.; Bhatt,~P.; Thomas,~A.; Ebbesen,~T.~W.; George,~J. Cavity catalysis
  by cooperative vibrational strong coupling of reactant and solvent molecules.
  \emph{Angew. Chem. Int. Ed.} \textbf{2019}, \emph{58}, 10635\relax
\mciteBstWouldAddEndPuncttrue
\mciteSetBstMidEndSepPunct{\mcitedefaultmidpunct}
{\mcitedefaultendpunct}{\mcitedefaultseppunct}\relax
\EndOfBibitem
\bibitem[Lather and George(2021)Lather, and George]{lather21}
Lather,~J.; George,~J. Improving enzyme catalytic efficiency by cooperative
  vibrational strong coupling of water. \emph{J. Phys. Chem. Lett.}
  \textbf{2021}, \emph{12}, 379--384\relax
\mciteBstWouldAddEndPuncttrue
\mciteSetBstMidEndSepPunct{\mcitedefaultmidpunct}
{\mcitedefaultendpunct}{\mcitedefaultseppunct}\relax
\EndOfBibitem
\bibitem[Hirai \latin{et~al.}(2020)Hirai, Hutchison, and Uji-i]{hirai20a}
Hirai,~K.; Hutchison,~J.~A.; Uji-i,~H. Recent progress of vibropolaritonic
  chemistry. \emph{ChemPlusChem} \textbf{2020}, \emph{85}, 1981--1988\relax
\mciteBstWouldAddEndPuncttrue
\mciteSetBstMidEndSepPunct{\mcitedefaultmidpunct}
{\mcitedefaultendpunct}{\mcitedefaultseppunct}\relax
\EndOfBibitem
\bibitem[Hirai \latin{et~al.}(2020)Hirai, Takeda, Hutchison, and
  Uji-i]{hirai20b}
Hirai,~K.; Takeda,~R.; Hutchison,~J.~A.; Uji-i,~H. Modulation of Prins
  cyclization by vibrational strong coupling. \emph{Angew. Chem. Int. Ed.}
  \textbf{2020}, \emph{59}, 5332--5335\relax
\mciteBstWouldAddEndPuncttrue
\mciteSetBstMidEndSepPunct{\mcitedefaultmidpunct}
{\mcitedefaultendpunct}{\mcitedefaultseppunct}\relax
\EndOfBibitem
\bibitem[Li \latin{et~al.}(2020)Li, Nitzan, and Subotnik]{li20a}
Li,~T.~E.; Nitzan,~A.; Subotnik,~J.~E. On the origin of ground-state
  vacuum-field catalysis: Equilibrium consideration. \emph{J. Chem. Phys.}
  \textbf{2020}, \emph{152}, 234107\relax
\mciteBstWouldAddEndPuncttrue
\mciteSetBstMidEndSepPunct{\mcitedefaultmidpunct}
{\mcitedefaultendpunct}{\mcitedefaultseppunct}\relax
\EndOfBibitem
\bibitem[Campos-Gonzalez-Angulo and Yuen-Zhou(2020)Campos-Gonzalez-Angulo, and
  Yuen-Zhou]{angulo20}
Campos-Gonzalez-Angulo,~J.~A.; Yuen-Zhou,~J. Polaritonic normal modes in
  transition state theory. \emph{J. Chem. Phys.} \textbf{2020}, \emph{152},
  161101\relax
\mciteBstWouldAddEndPuncttrue
\mciteSetBstMidEndSepPunct{\mcitedefaultmidpunct}
{\mcitedefaultendpunct}{\mcitedefaultseppunct}\relax
\EndOfBibitem
\bibitem[Galego \latin{et~al.}(2019)Galego, Climent, Garcia-Vidal, and
  Feist]{galego19}
Galego,~J.; Climent,~C.; Garcia-Vidal,~F.~J.; Feist,~J. Cavity Casimir-Polder
  Forces and Their Effects in Ground-State Chemical Reactivity. \emph{Phys.
  Rev. X} \textbf{2019}, \emph{9}, 021057\relax
\mciteBstWouldAddEndPuncttrue
\mciteSetBstMidEndSepPunct{\mcitedefaultmidpunct}
{\mcitedefaultendpunct}{\mcitedefaultseppunct}\relax
\EndOfBibitem
\bibitem[Climent \latin{et~al.}(2019)Climent, Galego, Garcia-Vidal, and
  Feist]{climent19}
Climent,~C.; Galego,~J.; Garcia-Vidal,~F.~J.; Feist,~J. Plasmonic Nanocavities
  Enable Self-Induced Electrostatic Catalysis. \emph{Angew. Chem., Int. Ed.}
  \textbf{2019}, \emph{58}, 8698--8702\relax
\mciteBstWouldAddEndPuncttrue
\mciteSetBstMidEndSepPunct{\mcitedefaultmidpunct}
{\mcitedefaultendpunct}{\mcitedefaultseppunct}\relax
\EndOfBibitem
\bibitem[Triana \latin{et~al.}(2020)Triana, Hern\'andez, and Herrera]{triana20}
Triana,~J.~F.; Hern\'andez,~F.~J.; Herrera,~F. The shape of the electric dipole
  function determines the sub-picosecond dynamics of anharmonic vibrational
  polaritons. \emph{J. Chem. Phys.} \textbf{2020}, \emph{152}, 234111\relax
\mciteBstWouldAddEndPuncttrue
\mciteSetBstMidEndSepPunct{\mcitedefaultmidpunct}
{\mcitedefaultendpunct}{\mcitedefaultseppunct}\relax
\EndOfBibitem
\bibitem[Vurgaftman \latin{et~al.}(2020)Vurgaftman, Simpkins, Dunkelberger, and
  Owrutsky]{vurgaftman20}
Vurgaftman,~I.; Simpkins,~B.~S.; Dunkelberger,~A.~D.; Owrutsky,~J.~C.
  Negligible effect of vibrational polaritons on chemical reaction rates via
  the density of states pathway. \emph{J. Phys. Chem. Lett.} \textbf{2020},
  \emph{11}, 3557--3562\relax
\mciteBstWouldAddEndPuncttrue
\mciteSetBstMidEndSepPunct{\mcitedefaultmidpunct}
{\mcitedefaultendpunct}{\mcitedefaultseppunct}\relax
\EndOfBibitem
\bibitem[Sch{\"a}fer \latin{et~al.}(2021)Sch{\"a}fer, Flick, Ronca, Narang, and
  Rubio]{schafer21}
Sch{\"a}fer,~C.; Flick,~J.; Ronca,~E.; Narang,~P.; Rubio,~A. Shining Light on
  the Microscopic Resonant Mechanism Responsible for Cavity-Mediated Chemical
  Reactivity. \emph{arXiv preprint arXiv:2104.12429} \textbf{2021}, \relax
\mciteBstWouldAddEndPunctfalse
\mciteSetBstMidEndSepPunct{\mcitedefaultmidpunct}
{}{\mcitedefaultseppunct}\relax
\EndOfBibitem
\bibitem[Flick \latin{et~al.}(2017)Flick, Ruggenthaler, Appel, and
  Rubio]{flick17a}
Flick,~J.; Ruggenthaler,~M.; Appel,~H.; Rubio,~A. Atoms and molecules in
  cavities, from weak to strong coupling in quantum-electrodynamics (qed)
  chemistry,. \emph{Proc. Natl. Acad. Sci. U. S. A.} \textbf{2017}, \emph{114},
  3026--3034\relax
\mciteBstWouldAddEndPuncttrue
\mciteSetBstMidEndSepPunct{\mcitedefaultmidpunct}
{\mcitedefaultendpunct}{\mcitedefaultseppunct}\relax
\EndOfBibitem
\bibitem[Power and Zienau(1959)Power, and Zienau]{power59}
Power,~E.~A.; Zienau,~S. Coulomb gauge in non-relativistic quantum
  electro-dynamics and the shape of spectral lines. \emph{Phil. Trans. R. Soc.
  Lond. A} \textbf{1959}, \emph{251}, 427\relax
\mciteBstWouldAddEndPuncttrue
\mciteSetBstMidEndSepPunct{\mcitedefaultmidpunct}
{\mcitedefaultendpunct}{\mcitedefaultseppunct}\relax
\EndOfBibitem
\bibitem[Zwanzig(1973)]{zwanzig73}
Zwanzig,~R. Nonlinear generalized Langevin equations. \emph{J. Stat. Phys.}
  \textbf{1973}, \emph{9}, 215--220\relax
\mciteBstWouldAddEndPuncttrue
\mciteSetBstMidEndSepPunct{\mcitedefaultmidpunct}
{\mcitedefaultendpunct}{\mcitedefaultseppunct}\relax
\EndOfBibitem
\bibitem[Fischer and Saalfrank(2021)Fischer, and Saalfrank]{fischer21}
Fischer,~E.~W.; Saalfrank,~P. Ground state properties and infrared spectra of
  anharmonic vibrational polaritons of small molecules in cavities. \emph{J.
  Chem. Phys.} \textbf{2021}, \emph{154}, 104311\relax
\mciteBstWouldAddEndPuncttrue
\mciteSetBstMidEndSepPunct{\mcitedefaultmidpunct}
{\mcitedefaultendpunct}{\mcitedefaultseppunct}\relax
\EndOfBibitem
\bibitem[Feynman and Vernon(2000)Feynman, and Vernon]{feynman00}
Feynman,~R.~P.; Vernon,~F.~L. The theory of a general quantum system
  interacting with a linear dissipative system. \emph{Ann. Phys.}
  \textbf{2000}, \emph{281}, 547--607\relax
\mciteBstWouldAddEndPuncttrue
\mciteSetBstMidEndSepPunct{\mcitedefaultmidpunct}
{\mcitedefaultendpunct}{\mcitedefaultseppunct}\relax
\EndOfBibitem
\bibitem[Hanggi \latin{et~al.}(1990)Hanggi, Talkner, and Borkovec]{hanggi90}
Hanggi,~P.; Talkner,~P.; Borkovec,~M. Reaction-rate theory: Fifty years after
  Kramers. \emph{Rev. Mod. Phys.} \textbf{1990}, \emph{62}, 251\relax
\mciteBstWouldAddEndPuncttrue
\mciteSetBstMidEndSepPunct{\mcitedefaultmidpunct}
{\mcitedefaultendpunct}{\mcitedefaultseppunct}\relax
\EndOfBibitem
\bibitem[Cao and Voth(1996)Cao, and Voth]{cao27}
Cao,~J.; Voth,~G.~A. A unified framework for quantum activated rate processes:
  I. General theory. \emph{J. Chem. Phys.} \textbf{1996}, \emph{105},
  6856\relax
\mciteBstWouldAddEndPuncttrue
\mciteSetBstMidEndSepPunct{\mcitedefaultmidpunct}
{\mcitedefaultendpunct}{\mcitedefaultseppunct}\relax
\EndOfBibitem
\bibitem[Wolynes(1981)]{wolynes81}
Wolynes,~P.~G. Quantum theory of activated events in condensed phases.
  \emph{Phys. Rev. Lett.} \textbf{1981}, \emph{47}, 968\relax
\mciteBstWouldAddEndPuncttrue
\mciteSetBstMidEndSepPunct{\mcitedefaultmidpunct}
{\mcitedefaultendpunct}{\mcitedefaultseppunct}\relax
\EndOfBibitem
\bibitem[Pollak(1986)]{pollak86a}
Pollak,~E. Transition state theory for quantum decay rates in dissipative
  systems: the high-temperature limit. \emph{Chem. Phys. Lett.} \textbf{1986},
  \emph{85}, 178\relax
\mciteBstWouldAddEndPuncttrue
\mciteSetBstMidEndSepPunct{\mcitedefaultmidpunct}
{\mcitedefaultendpunct}{\mcitedefaultseppunct}\relax
\EndOfBibitem
\bibitem[Grote and Hynes(1980)Grote, and Hynes]{grote80}
Grote,~R.~F.; Hynes,~J.~T. The stable states picture of chemical reactions. II.
  rate constants for condensed and gas phase reaction models. \emph{J. Chem.
  Phys.} \textbf{1980}, \emph{73}, 2715\relax
\mciteBstWouldAddEndPuncttrue
\mciteSetBstMidEndSepPunct{\mcitedefaultmidpunct}
{\mcitedefaultendpunct}{\mcitedefaultseppunct}\relax
\EndOfBibitem
\bibitem[Truhlar and Garrett(2000)Truhlar, and Garrett]{truhlar00}
Truhlar,~D.~G.; Garrett,~B.~C. Multidimensional transition state theory and the
  validity of Grote-Hynes theory. \emph{J. Phys. Chem. B} \textbf{2000},
  \emph{104}, 1069\relax
\mciteBstWouldAddEndPuncttrue
\mciteSetBstMidEndSepPunct{\mcitedefaultmidpunct}
{\mcitedefaultendpunct}{\mcitedefaultseppunct}\relax
\EndOfBibitem
\bibitem[Nitzan(2007)]{nitzan06}
Nitzan,~A. \emph{Chemical Dynamics in Condensed Phases: Relaxation, Transfer
  and Reactions in Condensed Molecular Systems}; Oxford University Press:
  Oxford, U.K., 2007\relax
\mciteBstWouldAddEndPuncttrue
\mciteSetBstMidEndSepPunct{\mcitedefaultmidpunct}
{\mcitedefaultendpunct}{\mcitedefaultseppunct}\relax
\EndOfBibitem
\bibitem[Cao and Voth(1996)Cao, and Voth]{cao26}
Cao,~J.; Voth,~G.~A. A theory for quantum activated rate constants in
  dissipative systems. \emph{Chem. Phys. Lett} \textbf{1996}, \emph{261},
  111\relax
\mciteBstWouldAddEndPuncttrue
\mciteSetBstMidEndSepPunct{\mcitedefaultmidpunct}
{\mcitedefaultendpunct}{\mcitedefaultseppunct}\relax
\EndOfBibitem
\bibitem[Pollak(1986)]{pollak86}
Pollak,~E. Theory of activated rate processes: A new derivation of Kramers'
  expression. \emph{J. Chem. Phys.} \textbf{1986}, \emph{85}, 865\relax
\mciteBstWouldAddEndPuncttrue
\mciteSetBstMidEndSepPunct{\mcitedefaultmidpunct}
{\mcitedefaultendpunct}{\mcitedefaultseppunct}\relax
\EndOfBibitem
\bibitem[Pollak \latin{et~al.}(1990)Pollak, Tucker, and Berne]{pollak90}
Pollak,~E.; Tucker,~S.~C.; Berne,~B.~J. Variational transition-state theory for
  reaction rates in dissipative systems. \emph{Phys. Rev. Lett.} \textbf{1990},
  \emph{65}, 1399\relax
\mciteBstWouldAddEndPuncttrue
\mciteSetBstMidEndSepPunct{\mcitedefaultmidpunct}
{\mcitedefaultendpunct}{\mcitedefaultseppunct}\relax
\EndOfBibitem
\bibitem[Li \latin{et~al.}(2021)Li, Mandal, and Huo]{li21a}
Li,~X.; Mandal,~A.; Huo,~P. Cavity frequency-dependent theory for vibrational
  polariton chemistry. \emph{Nat. Commun.} \textbf{2021}, \emph{12}, 1--9\relax
\mciteBstWouldAddEndPuncttrue
\mciteSetBstMidEndSepPunct{\mcitedefaultmidpunct}
{\mcitedefaultendpunct}{\mcitedefaultseppunct}\relax
\EndOfBibitem
\bibitem[Kryvohuz and Cao(2010)Kryvohuz, and Cao]{cao109}
Kryvohuz,~M.; Cao,~J. Noise-induced dynamics symmetry breaking and stochastic
  transitions in ABA molecules: II. Symmetric-antisymmetic normal mode
  switching. \emph{Chem. Phys.} \textbf{2010}, \emph{390}, 258\relax
\mciteBstWouldAddEndPuncttrue
\mciteSetBstMidEndSepPunct{\mcitedefaultmidpunct}
{\mcitedefaultendpunct}{\mcitedefaultseppunct}\relax
\EndOfBibitem
\bibitem[Gu and Mukamel(2020)Gu, and Mukamel]{gu20}
Gu,~B.; Mukamel,~S. Cooperative conical intersection dynamics of two pyrazine
  molecules in an optical cavity. \emph{J. Phys. Chem. Lett.} \textbf{2020},
  \emph{11}, 5555\relax
\mciteBstWouldAddEndPuncttrue
\mciteSetBstMidEndSepPunct{\mcitedefaultmidpunct}
{\mcitedefaultendpunct}{\mcitedefaultseppunct}\relax
\EndOfBibitem
\bibitem[Campos-Gonzalez-Angulo \latin{et~al.}(2019)Campos-Gonzalez-Angulo,
  Ribeiro, and Yuen-Zhou]{angulo19}
Campos-Gonzalez-Angulo,~J.~A.; Ribeiro,~R.~F.; Yuen-Zhou,~J. Resonant catalysis
  of thermally activated chemical reactions with vibrational polaritons.
  \emph{Nat. Commun.} \textbf{2019}, \emph{10}, 4685\relax
\mciteBstWouldAddEndPuncttrue
\mciteSetBstMidEndSepPunct{\mcitedefaultmidpunct}
{\mcitedefaultendpunct}{\mcitedefaultseppunct}\relax
\EndOfBibitem
\bibitem[Galego \latin{et~al.}(2017)Galego, Garcia-Vidal, and Feist]{galego17}
Galego,~J.; Garcia-Vidal,~F.~J.; Feist,~J. Many molecule reaction triggered by
  a single photon in polaritonic chemistry. \emph{Phys. Rev. Lett.}
  \textbf{2017}, \emph{119}, 136001\relax
\mciteBstWouldAddEndPuncttrue
\mciteSetBstMidEndSepPunct{\mcitedefaultmidpunct}
{\mcitedefaultendpunct}{\mcitedefaultseppunct}\relax
\EndOfBibitem
\bibitem[Phuc \latin{et~al.}(2020)Phuc, Ishizaki, and Trung]{phuc20}
Phuc,~N.~T.; Ishizaki,~A.; Trung,~P.~Q. Controlling the electron-transfer
  reaction rate through molecular-vibration polaritons in the ultrastrong
  coupling regime. \emph{Sci. Rep.} \textbf{2020}, \emph{10}, 7318\relax
\mciteBstWouldAddEndPuncttrue
\mciteSetBstMidEndSepPunct{\mcitedefaultmidpunct}
{\mcitedefaultendpunct}{\mcitedefaultseppunct}\relax
\EndOfBibitem
\bibitem[Moix \latin{et~al.}(2012)Moix, Zhao, and Cao]{cao121}
Moix,~J.; Zhao,~Y.; Cao,~J. Equilibrium-reduced density matrix formulation:
  Influence of noise, disorder, and temperature on localization in excitonic
  systems. \emph{Phys. Rev. B} \textbf{2012}, \emph{85}, 115412\relax
\mciteBstWouldAddEndPuncttrue
\mciteSetBstMidEndSepPunct{\mcitedefaultmidpunct}
{\mcitedefaultendpunct}{\mcitedefaultseppunct}\relax
\EndOfBibitem
\bibitem[Herrera and Spano(2016)Herrera, and Spano]{herrera16}
Herrera,~F.; Spano,~F.~C. Cavity-controlled chemistry in molecular ensembles.
  \emph{Phys. Rev. Lett.} \textbf{2016}, \emph{116}, 238301\relax
\mciteBstWouldAddEndPuncttrue
\mciteSetBstMidEndSepPunct{\mcitedefaultmidpunct}
{\mcitedefaultendpunct}{\mcitedefaultseppunct}\relax
\EndOfBibitem
\bibitem[Scholes \latin{et~al.}(2020)Scholes, DelPo, and Kudisch]{scholes20}
Scholes,~G.~D.; DelPo,~C.~A.; Kudisch,~B. Entropy reorders polariton states.
  \emph{J. Phys. Chem. Lett.} \textbf{2020}, \emph{T11}, 6389\relax
\mciteBstWouldAddEndPuncttrue
\mciteSetBstMidEndSepPunct{\mcitedefaultmidpunct}
{\mcitedefaultendpunct}{\mcitedefaultseppunct}\relax
\EndOfBibitem
\bibitem[Lee \latin{et~al.}(2015)Lee, Moix, and Cao]{cao150}
Lee,~C.~K.; Moix,~J.~M.; Cao,~J. Coherent quantum transport in disordered
  systems: A unified polaron treatment of hopping and band-like transport.
  \emph{J. Chem. Phys.} \textbf{2015}, \emph{142}, 164103\relax
\mciteBstWouldAddEndPuncttrue
\mciteSetBstMidEndSepPunct{\mcitedefaultmidpunct}
{\mcitedefaultendpunct}{\mcitedefaultseppunct}\relax
\EndOfBibitem
\bibitem[Wers\"all \latin{et~al.}(2019)Wers\"all, Munkhbat, Baranov, Herrera,
  Cao, Antosiewicz, and Shegai]{cao192}
Wers\"all,~M.; Munkhbat,~B.; Baranov,~D.; Herrera,~F.; Cao,~J.;
  Antosiewicz,~T.~J.; Shegai,~T. Correlative Dark-Field and Photoluminescence
  Spectroscopy of Individual Plasmon-Molecule Hybrid Nanostructures in a Strong
  Coupling Regime. \emph{ACS Photonics} \textbf{2019}, \emph{6},
  2570--2576\relax
\mciteBstWouldAddEndPuncttrue
\mciteSetBstMidEndSepPunct{\mcitedefaultmidpunct}
{\mcitedefaultendpunct}{\mcitedefaultseppunct}\relax
\EndOfBibitem
\bibitem[Voth \latin{et~al.}(1989)Voth, Chandler, and Miller]{voth89}
Voth,~G.~A.; Chandler,~D.; Miller,~W.~H. Rigorous formulation of quantum
  transition state theory and its dynamical corrections. \emph{J. Chem. Phys.}
  \textbf{1989}, \emph{91}, 7749--7760\relax
\mciteBstWouldAddEndPuncttrue
\mciteSetBstMidEndSepPunct{\mcitedefaultmidpunct}
{\mcitedefaultendpunct}{\mcitedefaultseppunct}\relax
\EndOfBibitem
\bibitem[Cao and Voth(1995)Cao, and Voth]{cao19}
Cao,~J.; Voth,~G.~A. Modeling physical systems by effective harmonic
  oscillators: The optimized quadratic approximation. \emph{J. Chem. Phys.}
  \textbf{1995}, \emph{102}, 3337\relax
\mciteBstWouldAddEndPuncttrue
\mciteSetBstMidEndSepPunct{\mcitedefaultmidpunct}
{\mcitedefaultendpunct}{\mcitedefaultseppunct}\relax
\EndOfBibitem
\bibitem[Miller(1975)]{miller75}
Miller,~W.~H. Semiclassical limit of quantum mechanical transition state theory
  for non-separable systems. \emph{J. Chem. Phys.} \textbf{1975}, \emph{62},
  1899--1906\relax
\mciteBstWouldAddEndPuncttrue
\mciteSetBstMidEndSepPunct{\mcitedefaultmidpunct}
{\mcitedefaultendpunct}{\mcitedefaultseppunct}\relax
\EndOfBibitem
\bibitem[Cao \latin{et~al.}(1995)Cao, Minichino, and Voth]{cao21}
Cao,~J.; Minichino,~C.; Voth,~G.~A. The computation of electron transfer rates:
  The nonadiabatic instanton solution. \emph{J. Chem. Phys.} \textbf{1995},
  \emph{103}, 1391\relax
\mciteBstWouldAddEndPuncttrue
\mciteSetBstMidEndSepPunct{\mcitedefaultmidpunct}
{\mcitedefaultendpunct}{\mcitedefaultseppunct}\relax
\EndOfBibitem
\bibitem[Garcia-Vidal \latin{et~al.}(2021)Garcia-Vidal, Ciuti, and
  Ebbesen]{garcia21}
Garcia-Vidal,~F.; Ciuti,~C.; Ebbesen,~T. Manipulating matter by strong coupling
  to vacuum fields. \emph{Science} \textbf{2021}, \emph{373}, 6551\relax
\mciteBstWouldAddEndPuncttrue
\mciteSetBstMidEndSepPunct{\mcitedefaultmidpunct}
{\mcitedefaultendpunct}{\mcitedefaultseppunct}\relax
\EndOfBibitem
\end{mcitethebibliography}

\end{document}


\title{-SUPPORTING INFORMATION- \\
for \\ Quantum Effects in Chemical Reactions under Polaritonic Vibrational Strong Coupling}
\author{Pei-Yun Yang}
\affiliation{Department of Chemistry, Massachusetts Institute of Technology, Massachusetts, 02139 USA }
\author{Jianshu Cao}
\affiliation{Department of Chemistry, Massachusetts Institute of Technology, Massachusetts, 02139 USA }

\maketitle
\renewcommand{\thepage}{S\arabic{page}}
\renewcommand{\thesection}{S\arabic{section}}
\renewcommand{\theequation}{S\arabic{equation}}
\renewcommand{\thefigure}{S\arabic{figure}}

\newpage

\section{Perturbation Analysis: Single Molecule VSC}

\subsection{Frequency sum in the reactant well}
\label{sec1.1}
In general, direct perturbation expansion of eigen frequency diverges at resonance, as shown in the next subsection. Since the zero-point-energy (ZPE)
shift involves
the sum of the eigen frequencies, we can obtain an exact expression of the frequency sum and then evaluate it with perturbation expansion.
This approach can be compared with the direct perturbation of eigen frequency in the next subsection.
The Hessian matrix of a single molecule in the VSC regime is given in the main text and is rewritten here as
\begin{align}
A_{2}=\left[
\begin{array}{cc}
\omega _{e}^{2} & J^{2} \\
J^{2} & \omega _{c}^{2}
\end{array}
\right]
\label{A2}
\end{align}
where the two new variables are introduced as
\begin{subequations}
\begin{align}
J^{2} =~&g\omega _{c}^{2} \\
\omega _{e}^{2} =~&\omega ^{2}+g^{2}\omega _{c}^{2}=\omega^{2}+J^{4}/\omega _{c}^{2}
\end{align}
\end{subequations}
The trace and determinant of $A_2$  are invariant and can be written explicitly as
\begin{subequations}
\begin{align}
{\rm Tr}\left[ A_{2}\right] =&\lambda _{+}^{2}+\lambda _{-}^{2}=\omega
_{e}^{2}+\omega _{c}^{2} \\
\det \left[ A_{2}\right] =&\lambda _{+}^{2}\lambda _{-}^{2}=\omega
_{e}^{2}\omega _{c}^{2}-J^{4}
\end{align}
\end{subequations}
Then, the frequency sum is expressed as
\begin{equation}
\lambda _{+}+\lambda _{-}
=\sqrt{\omega _{e}^{2}+\omega _{c}^{2}+2\sqrt{\omega _{e}^{2}\omega
_{c}^{2}-J^{4}}}
\end{equation}
Now taking advantage of the straightforward expansion
\begin{align}
\sqrt{a^{2}+\sqrt{b^{4}-4J^{4}}}
= \sqrt{a^{2}+b^{2}}-\frac{J^{4}}{b^{2}\sqrt{a^{2}+b^{2}}}+O\left( J^8 \right)
\end{align}
we arrive at
\begin{align}
\lambda _{+}+\lambda _{-}  &=~\omega _{e}+\omega _{c}-\frac{J^{4}}{\left( 2\omega _{e}\omega
_{c}\right) \left( \omega _{e}+\omega _{c}\right) }+O\left( J^8 \right) \notag\\
&=~\omega+\omega _{c}+\frac{J^{4}}{2\omega _{c}^{2}\left( \omega
+\omega _{c}\right) } +O\left( J^8 \right)  \notag\\
&=~\omega+\omega _{c}+\frac{g^{2}\omega _{c}^{2}}{2\left( \omega
+\omega _{c}\right) } +O\left( g^4 \right)
\label{sum}
\end{align}
Here, we also use the expansion of the effective frequency,
\begin{align}
\omega _{e} =~&\left[ \omega^{2}+J^{4}/\omega _{c}^{2}\right] ^{\frac{1%
}{2}} \notag\\
=~&\omega+\frac{J^{4}}{2\omega\omega _{c}^{2}}+O\left( J^8 \right)
\end{align}
Evidently, the ZPE frequency shift in the reactant well is always positive
\begin{align}
\frac{J^{4}}{2\omega _{c}^{2}\left( \omega+\omega _{c}\right) }>0
\end{align}

\subsection{Frequency perturbation in the reactant well}

To diagonalize the Hessian matrix $A_2$, we introduce new variables
\begin{subequations}
\begin{align}
\omega _{s}^{2} =~&\frac{1}{2}\left( \omega _{e}^{2}+\omega _{c}^{2}\right)
\\
\delta ^{2} =~&\frac{1}{2}\left( \omega _{e}^{2}-\omega _{c}^{2}\right) \\
J^{2} =~&g\omega _{c}^{2}
\end{align}
\end{subequations}
and then explicitly write the eigenvalues of $A_2$ as
\begin{align}
\lambda _{\pm }^{2}=\omega _{s}^{2}\pm \sqrt{\delta ^{4}+J^{4}}
\end{align}
Next, we evaluate the eigenfrequency perturbatively. giving
\begin{align}
\lambda _{\pm } =&\sqrt{\omega _{s}^{2}\pm \sqrt{\delta ^{4}+J^{4}}} \notag\\
=&\sqrt{\omega _{s}^{2}\pm \delta ^{2}}\pm \frac{J^{4}}{4\delta ^{2}\sqrt{%
\omega _{s}^{2}\pm \delta ^{2}}}+O\left( J\right) ^{8}
\end{align}
or, explicitly,
\begin{subequations}
\begin{align}
\lambda _{+} =~&\omega _{e}+\frac{1}{4\omega _{e}}\frac{J^{4}}{\delta ^{2}}%
+O\left( J^8 \right)  \\
\lambda _{-} =~&\omega _{c}-\frac{1}{4\omega _{c}}\frac{J^{4}}{\delta ^{2}}%
+O\left( J^8 \right)
\end{align}
\end{subequations}
The above perturbative expressions diverge at resonance, $\omega=\omega_c$ (i.e. $\delta=0$).
Interestingly, the divergencies of the two frequencies cancel at resonance, such that
the sum of the two eigen frequencies is well-behaved, giving,
\begin{align}
\lambda _{+}+\lambda _{-}
=~\omega _{e}+\omega _{c}-\frac{J^{4}}{2\omega _{e}\omega _{c}\left( \omega
_{e}+\omega _{c}\right) }+O\left( J^8 \right)
\end{align}
which agrees with Eq.~(\ref{sum})

\subsection{Frequency perturbation at the reactive barrier}

The analysis of the transition state  (TS)  or the reactant barrier
follows closely the previous subsection on the reactant well except for the change from the vibrational frequency
$\omega$ to the barrier frequency $\omega_{\ddagger}$.
The Hessian matrix at TS is given as
\begin{align}
A_{2b}=\left[
\begin{array}{cc}
-\omega _{e\ddagger}^{2} & J_\ddagger^{2} \\
J_\ddagger^{2} & \omega _{c}^{2}%
\end{array}%
\right]
\end{align}
where the two new parameters are defined as
\begin{subequations}
\begin{align}
J_\dd^2 & = g_{\dd}  \omega_c^2 \\
-\omega _{e\ddagger}^{2} & = -\omega _{\ddagger}^{2}+g_\ddagger^{2}\omega _{c}^{2} \notag \\
&= -\omega _{\ddagger}^{2}+J_\ddagger^{4}/\omega _{c}^{2}
\end{align}
\end{subequations}
The Hessian matrix  $A_{2b}$ can be solved to yield a pair of eigen-values,
\begin{align}
\pm\lambda _{\pm b}^{2}=\omega _{s}^{2}\pm \sqrt{\delta ^{4}+J_\ddagger^{4}}
\label{eigen}
\end{align}
where two new parameters are introduced,
\begin{align}
\omega _{s}^{2} =~&\frac{1}{2}\left( \omega _{c}^{2}-\omega _{e\ddagger}^{2}\right) \notag
\\
\delta ^{2} =~&\frac{1}{2}\left( \omega _{e\ddagger}^{2}+\omega _{c}^{2}\right) \notag
\end{align}
We then expand  Eq.~(\ref{eigen}) in terms of the VSC strength to yield,
\begin{align}
\lambda _{\pm b} =~&\sqrt{\pm\omega _{s}^{2}+ \sqrt{\delta ^{4}+J_\ddagger^{4}}} \notag\\
=&\sqrt{\delta ^{2}\pm\omega _{s}^{2}}+ \frac{J_\ddagger^{4}}{4\delta ^{2}\sqrt{%
\delta ^{2}\pm\omega _{s}^{2}}}+O\left( J^8_\ddagger\right)
\end{align}
which leads to the stable eigen-frequency, $\lambda_{b}$, and the unstable eigen-frequency, $\lambda_{\dd}$,
 \begin{align}
\lambda _{+b}\equiv\lambda _{b} =~&\omega _{c}+\frac{J_\ddagger^{4}}{2\omega _{c}\left( \omega
_{e\ddagger}^{2}+\omega _{c}^{2}\right) }+O\left( J^8_\ddagger\right)  \notag\\
=~&\omega _{c}+\frac{J_\ddagger^{4}}{2\omega _{c}\left( \omega _{\ddagger}^{2}+\omega
_{c}^{2}\right) }+O\left( J^8_\ddagger\right)
\end{align}
\begin{align}\label{lambda}
\lambda _{-b}\equiv\lambda_\ddagger=\omega_{e\ddagger}+\frac{J_\ddagger^{4}}{2 \omega _{e\ddagger}\left( \omega
_{e\ddagger}^{2}+\omega _{c}^{2}\right)}+O\left( J^8_\ddagger\right)
\end{align}
Finally,  we arrive at the ZPE frequency shift
\begin{align}
S =~&\lambda _{+}+\lambda _{-}-\lambda _{b}-\omega \notag\\
\simeq~&\frac{1}{2\omega _{c}}\left[ \frac{J^{4}}{\omega \omega _{c}+\omega
_{c}^{2}}-\frac{J_\ddagger^{4}}{\omega _{\ddagger}^{2}+\omega _{c}^{2}}\right]  \notag\\
\simeq~&\frac{\omega _{c}}{2}\left[ \frac{g^{2}}{\omega \omega _{c}+\omega
_{c}^{2}}-\frac{g_\ddagger^{2}}{\omega _{\ddagger}^{2}+\omega _{c}^{2}}\right] \label{shift}
\end{align}
which is a key result of the perturbation analysis and is discussed extensively in the main text.

\subsection{Numerical examples}

\begin{figure}
\centerline{\scalebox{0.5}{\includegraphics[trim=0 0 0 0,clip]{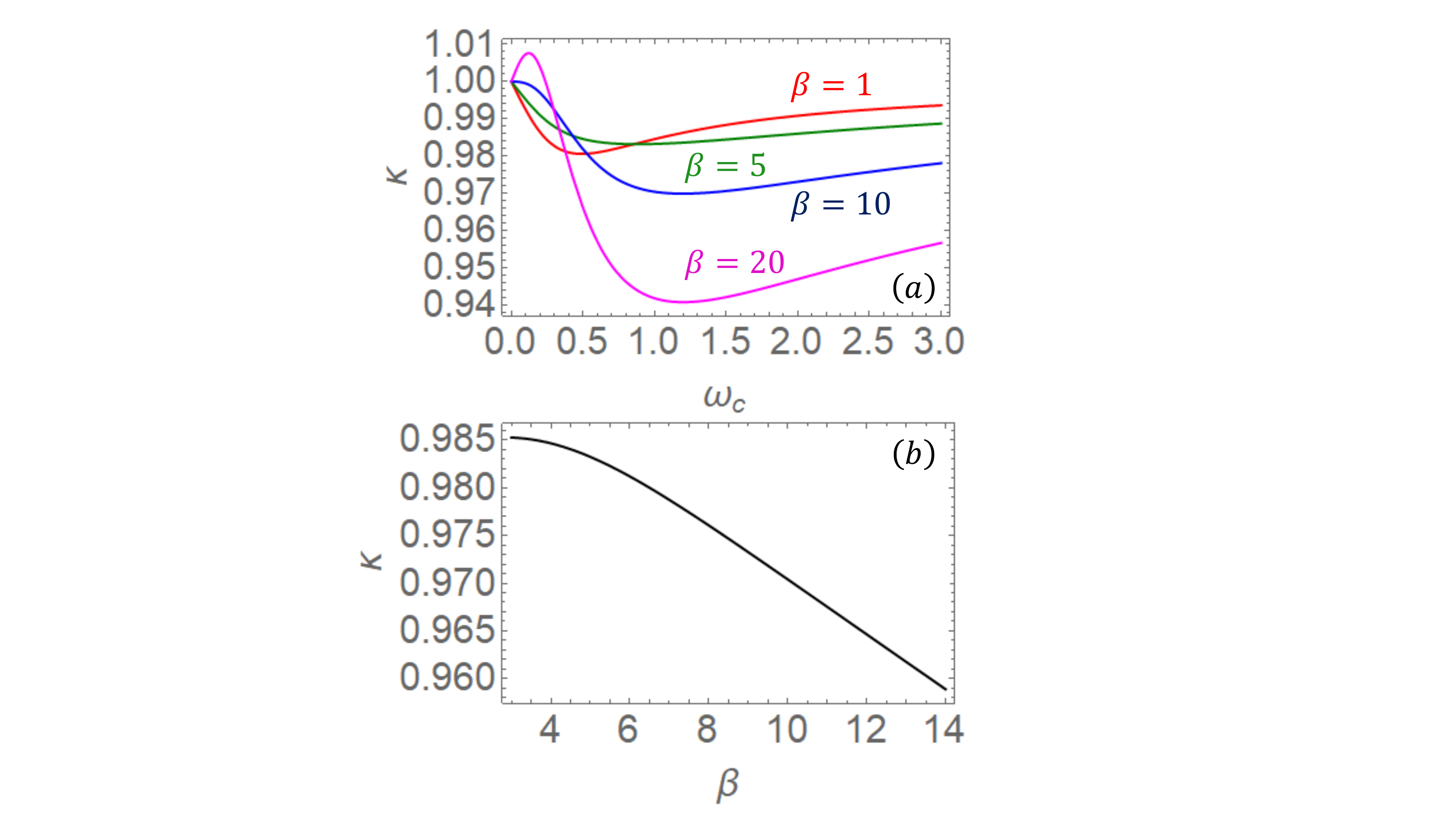}}}
\caption{(a) $\kappa$ as a function of $\omega_c$ for different inverse temperature $\beta$.
(b) $\kappa$  as a function of $\beta$ at $\omega_c=1$.
In this figure, all physical quantitis are in unit of $\omega$,  and relevant parameters are
 $\eta=\eta_{\ddagger}=0.1$ and $\omega_{\ddagger}=0.5$. }
\label{Fig1}
\end{figure}

First,  to examine the temperature dependence,  in Fig.~S1,
we plot $\kappa$ as a function of $\omega_c$  at inverse temperature $\beta$  (upper panel) and then $\kappa$  as a function of $\beta$
at the fixed cavity frequency of $\omega_c=1$ (lower panel).
Evidently, at temperature decreases, the minimal correction factor changes from the barrier
resonance to vibrational resonance, and the resonant amplitude increases drastically.  This figure supplements  Fig.~S1 in the main text.

\begin{figure}
\centerline{\scalebox{0.5}{\includegraphics[trim=0 0 0 0,clip]{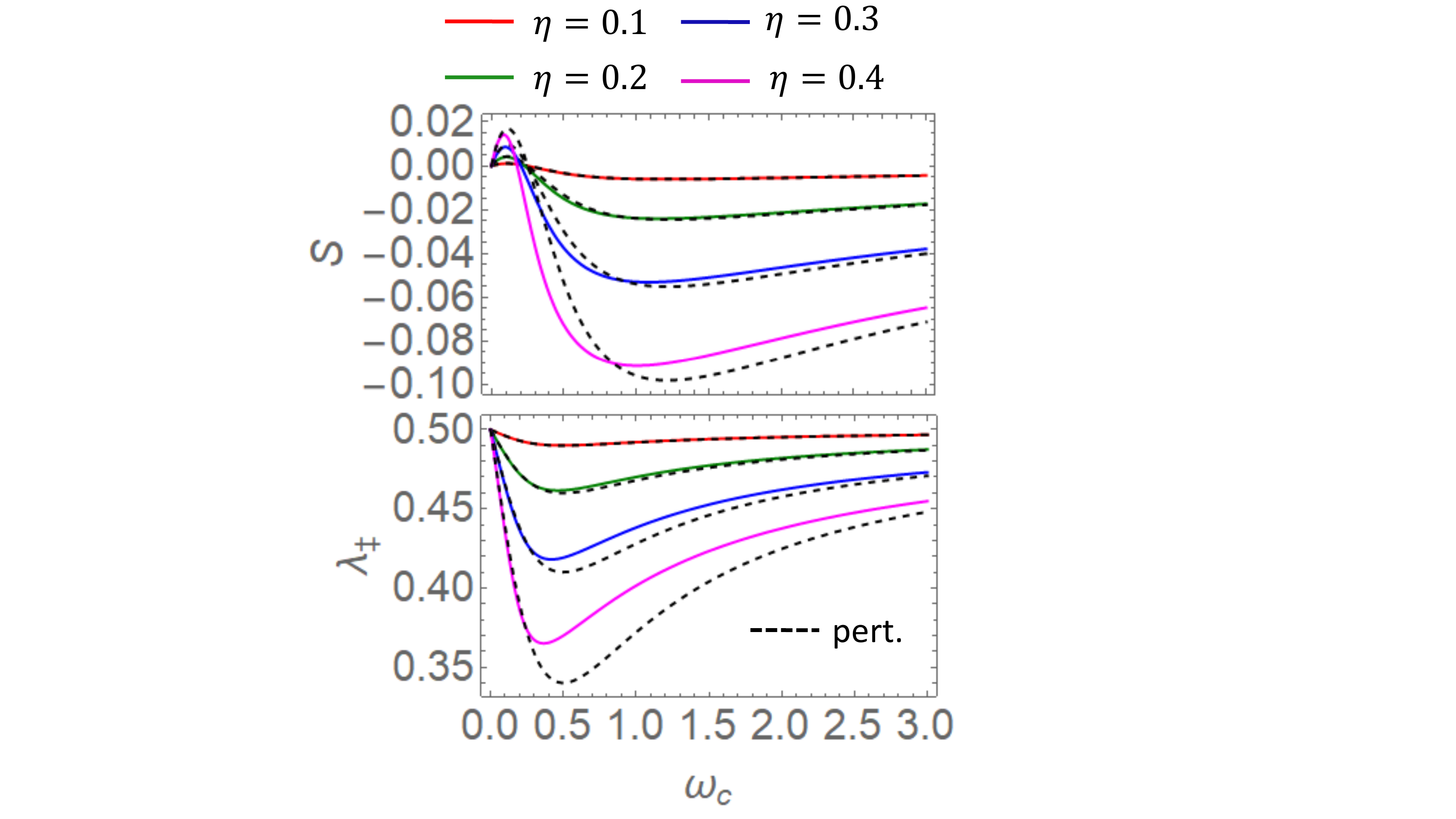}}}
\caption{The ZPE frequency shift $S$  (upper panel) and the unstable eigen frequency $\lambda_\ddagger$ (lower panel)
as a function of $\omega_c$  for different values of VSC strength $\eta=\eta_{\ddagger}$ at the fixed barrier frequency
 $\omega_{\ddagger}=0.5$.  The perturbative results are shown as the dashed line. All physical quantities are in unit of $\omega$.}
\label{Fig2}
\end{figure}

Next, to verify the perturbation expansion,  we calibrate the ZPE frequency shift $S$  (upper panel)
and the unstable eigen frequency $\lambda_\ddagger$ for different values of VSC strength $\eta=\eta_{\ddagger}$ .
As shown in Fig.~S2, the agreement is nearly perfect in the strong coupling regime ($\eta \le 0.1$) and exhibits
difference in the ultra-strong coupling regime ($\eta > 0.1$).  Interestingly,   the minimal of
the ZPE frequency shift $S$  occurs near the vibrational resonance $\omega_c \approx \omega$ whereas
the minimal of the unstable eigen frequency $\lambda_\ddagger$ occurs near the barrier resonance $\omega_c \approx \omega_\dd$.
This figure supplements  Fig.~2 in the main text.

\begin{figure}
\centerline{\scalebox{0.5}{\includegraphics[trim=0 0 0 0,clip]{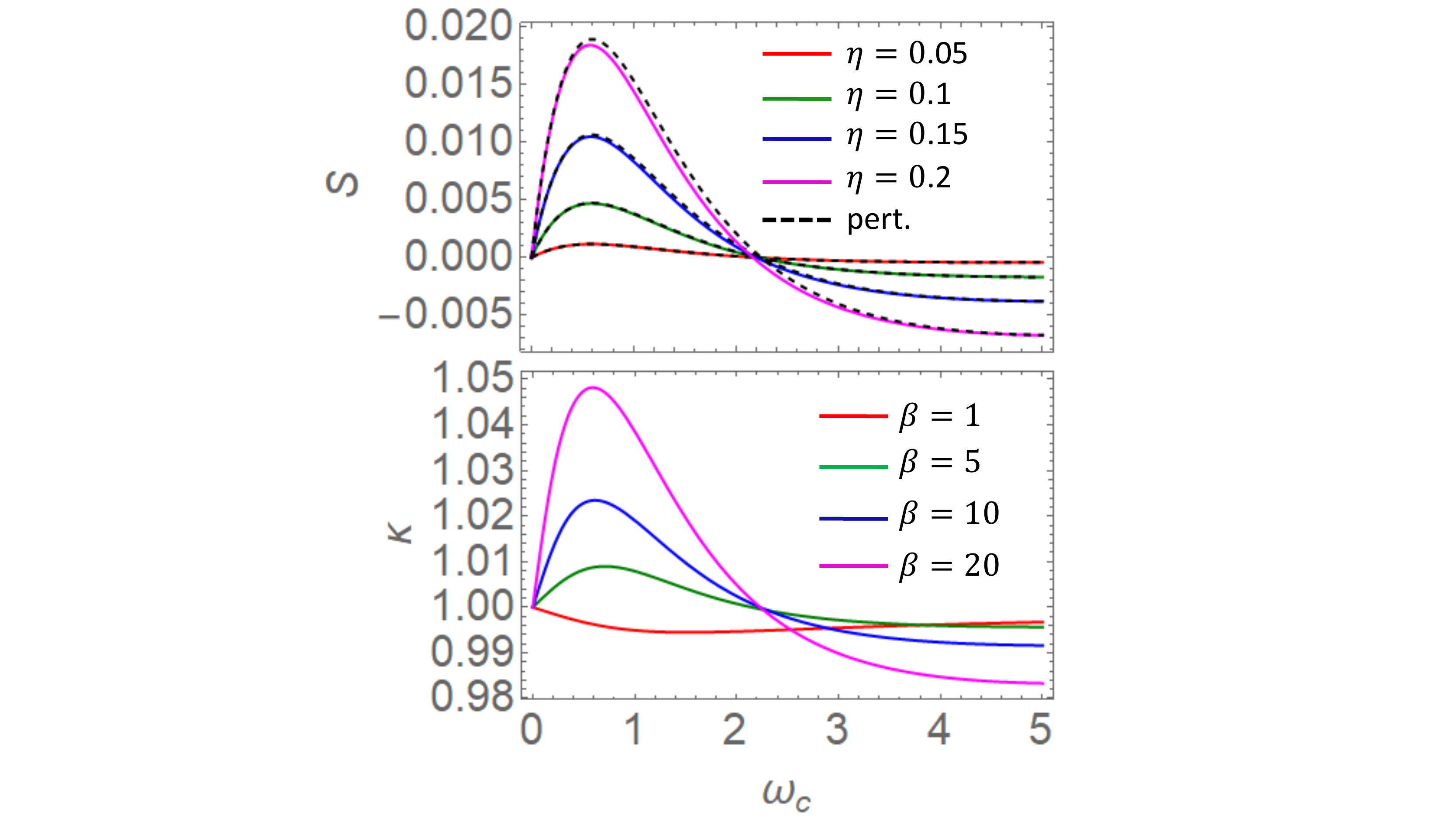}}}
\caption{ $S$  and $\kappa$ as a function of $\omega_c$ for $\omega_{\ddagger}=1.5$.
The  upper panel is the ZPE frequency shift $S$  for  different values of the VSC strength $\eta=\eta_{\ddagger}$.
The dashed lines are the perturbation results for comparison.
The lower panel is the the cavity-induced correction $\kappa$ at different inverse temperatures $\beta$
with parameters $\eta=\eta_{\ddagger}=0.1$.  All physical quantities are in unit of $\omega$.}
\label{Fig3}
\end{figure}

Finally, to examine the dependence on the barrier frequency $\omega_{\dd}$,  Fig.~S3 shows the results for $\omega_{\ddagger}=1.5$, larger than the vibrational frequency, whereas all the other calculations adopts $\omega_\dd=0.5$, smaller than the vibrational frequency.
The  upper panel is the ZPE frequency shift $S$  for  different values of the VSC strength $\eta=\eta_{\ddagger}$.
For comparison,  perturbation results show excellent agreement.
The lower panel is the the cavity-induced correction $\kappa$ at different inverse temperatures $\beta$
for a fixed VSC strength $\eta=\eta_{\ddagger}=0.1$.  In both panels, $\kappa>1$  except for a large cavity frequency,
indicating cavity-induced enhancement in reactivity.  In both panels, the peak enhancement occurs near the vibrational resonance $\omega_c=1.0$.
 The difference between $\omega_\dd=0.5$ and $\omega_\dd=1.5$ is consistent with the perturbation analysis.

\subsection{Minimal in the unstable frequency}

As discussed in the main text, the cavity-induced correction to the quantum TST rate, $\kappa$, is dominated by the frequency shift of
Eq.~(\ref{shift}) in the quantum regime,  which is experimentally relevant.  Theoretically, at high temperature $\hbar\beta\omega < 1$ (which is experimentally unrealistic),  the correction factor can be approximated by the Grote-Hynes (GH) correction, i.e.,
\begin{equation}
\kappa \approx \kappa_{GH} = { \lambda_\dd \over \omega_\dd }
\end{equation}
where $\lambda_{\dd}$ is given in Eq.~(\ref{lambda}).
Here we first establish the minimal of $\lambda_{\dd}$,  in agreement with a recent study [Nature Communications 12, p1 (2021)],  and
then use perturbation expansion to explain its physical origin.

\subsubsection{Exact solution}

The eigen-frequency of the unstable mode, as given formally in Eq.~(\ref{eigen}), is rewritten explicitly as
\begin{align}
\lambda^2_\ddagger = \frac{1}{2} [\Omega^4 + 4\omega^2_\ddagger \omega^2_{c}]^{1/2} - \frac{\Omega^2}{2}
\end{align}
where $\Omega$ is a frequency variable (not to be confused with the Rabi frequency, $\Omega_R$)
\begin{align}
	\Omega^2 = \omega^2_c - \omega^2_\ddagger + \omega^2_c g_\ddagger^2 = \omega^2_c - \omega^2_\ddagger +  B \omega_c
\end{align}
Here, we introduce a new parameter  via the following relation
\begin{equation}
g_\ddagger^2\omega^2_c = 4 \eta_\ddagger \omega \omega_c =  B\omega_c
\label{B}
\end{equation}
To find the minimal in $\lambda_\ddagger$, we take derivative with respect to
$\omega_c$ and obtain
\begin{align}
	\frac{d\lambda^2_\ddagger}{d\omega_c} =
	\frac{1}{2}\left[\frac{\Omega^2 \frac{d\Omega^2}{d\omega_c} + 4\omega^2_\ddagger\omega_c}
		 {(\Omega^4 + 4\omega^2_\ddagger\omega^2_c)^{1/2} }
		 -(2\omega_c +B)\right] = 0
\end{align}
where ${ d\Omega^2 / d\omega_c} = 2\omega_c + B$ .
Then, we find the condition for the optimal cavity frequency $\omega_c$,
\begin{align}	
	\Omega^2 (2\omega_c +B) + 4\omega^2_\ddagger \omega_c = (2\omega_c+B)(\Omega^4 + 4\omega^2_\ddagger\omega^2_c)^{1/2} .
\end{align}
Taking square on both sides and simplifying the expression, we arrive at
\begin{equation}
	2\omega^2_c + B \omega_c -2\omega^2_\ddagger = 0
\end{equation}
which predicts $\omega_c \approx \omega_\dd$ as $\eta_\dd < 1$.

\subsubsection{Perturbation analysis}

For a general 2-mode model, the Hessian matrix is given as
\begin{align}
	A_2 &=\left[
	\begin{array}[c]{cc}
		\omega^2_1	+ g_1^2 \omega^2_c&	g_1\omega^2_c
	\\
		g_1\omega^2_c						&\omega^2_c
	\end{array}\right]
\end{align}
which is the same as Eq.~(\ref{A2}).
The eigenvalue associated with the reactant mode is given perturbatively as
\begin{equation}
	\lambda^2_1	=	\omega^2_1 +g_1^2 \omega^2_c +
					\frac{(g_1\omega^2_c)^2}{\omega^2_1 - \omega^2_c}
\end{equation}
At the reactive barrier (i.e., TS), we let  $\omega^2_1 = - \omega^2_\ddagger$ and $g_1=g_\ddagger$, and then have the unstable eigenvalue as
\begin{equation}
	\lambda^2_\ddagger  =	\omega^2_\ddagger -g^2_\ddagger \omega^2_c +\frac{(g_\ddagger\omega^2_c)^2}{\omega^2_\ddagger + \omega^2_c}
					 =	\omega^2_\ddagger - \frac{(g_\ddagger\omega_c)^2\omega^2_\ddagger}{\omega^2_\ddagger + \omega^2_c}
\end{equation}
Using the definition of $g_\dd$ in Eq.~(\ref{B}), we have the unstable eigenvalue
\begin{align}
	\lambda^2_\ddagger = \omega^2_\ddagger
	\underbrace{-\frac{B \omega_c}{\omega^2_\ddagger + \omega^2_c} }_{\text{maximum at }\omega_\ddagger = \omega_c}
	\omega^2_\ddagger
\end{align}
which has a minimal at $\omega_c= \omega_\ddagger $.  Evidently, the perturbation analysis identifies two reasons for the minimal $\kappa_{GH}$:
(1)  the competition between the DSE term and the linear dipole coupling term;  (2) the parametric dependence of $g_\dd$ on the cavity
frequency in Eq.~(\ref{B}).   However, we emphasize that under experimental conditions  the cavity-induced correction $\kappa$ is dominated by
the quantum ZPE shift and the classical Grote-Hynes factor is not relevant.

\section{Perturbation Analysis: N-Molecule VSC}

\subsection{Three-mode solution}

We begin with a general analysis of the three-mode model for an ensemble of N molecules in cavity, described by
\begin{align}
U_3 = \frac{1}{2}\omega_1^2 q_1^2+\frac{1}{2}\omega^2 q_{N-1}^2+\frac{1}{2}\omega_c^2\left[q_c +g_1 q_1 +g_{N-1} q_{N-1} \right]^2
\end{align}
where $q_1$ is the reactive mode  and $q_{N-1}$ is the collective bright mode for the N-1 molecules excluding the reactive molecule.
With the rest of the N-2 modes as the dark state,
 the 3-mode expression is equivalent to the N-molecule potential described in the main text.
Then, the Hessian matrix of the 3 mode model is given as
\begin{align}
	A_3 = \left[	
	\begin{array}{clr}
		\omega^2_1 + g_1^2 \omega^2_c	&	g_1 g_{N-1}\omega^2_c	&g_1\omega^2_c	\\
		g_1 g_{N-1}\omega^2_c				&	\omega^2 + g_{N-1}^2 \omega^2_c	&	g_{N-1}\omega^2_c\\
		g_1\omega^2_c					&	g_{N-1}\omega^2_c			&\omega^2_c
	\end{array}\right]
	\label{A3}
\end{align}
where $g_{N-1}= g_1 \sqrt{N-1} $ is the collective N-1 molecule coupling strength and $g_1=g$ is the coupling strength of the single reactive mode.
 The trace and determinant of $A_3$ matrix are
\begin{align}
{\rm Tr} \left[A\right]=~&\omega^2_1 + g_1^2 \omega^2_c+\omega^2_c+\omega^2+ g_{N-1}^2 \omega^2_c, \\
\det\left[A\right]=~&\omega^2_1\omega^2\omega^2_c
\end{align}
Perturbative evaluation of $A_3$ yields,
\begin{subequations}
\begin{align}
\lambda_1^2=~&\omega_1^2+\frac{\left(g_1g_{N-1}\omega_c^2\right)^2}{\omega_1^2-\omega^2}+\frac{\left(g_1\omega_1\omega_c\right)^2}{\omega_1^2-\omega_c^2}, \\ \lambda_c^2=~&\omega_c^2+\frac{\left(g_1\omega_c^2\right)^2}{\omega_c^2-\omega_1^2}+\frac{\left(g_{N-1}\omega_c^2\right)^2}{\omega_c^2-\omega^2}, \\
\lambda_{N-1}^2=~&\omega^2+\frac{\left(g_{N-1}\omega\omega_c\right)^2}{\omega^2-\omega_c^2}+\frac{\left(g_{N-1}g_1\omega_c^2\right)^2}{\omega^2-\omega_1^2},
\end{align}
\end{subequations}
where $\lambda_1$ is the eigenvalue associated with the reactive mode,  $\lambda_c$ is the eigenvalue associated with the cavity mode,
and $\lambda_{N-1}$ is the eigenvalue associate with the collective bright state.
Taking the square root of the above eigenvalues and keeping the leading order of $g_1$ or $g_{N-1}$,  we have
\begin{subequations}
\begin{align}
\lambda_1=~&\omega_1+\frac{\omega_c^2}{2\omega_1}\frac{\left(g_1\omega_1\right)^2}{\omega_1^2-\omega_c^2}, \\
\lambda_c=~&\omega_c+\frac{\omega_c^2}{2\omega_c}\left[\frac{\left(g_1\omega_c\right)^2}{\omega_c^2-\omega_1^2}+\frac{\left(g_{N-1}\omega_c\right)^2}{\omega_c^2-\omega^2}\right], \\
\lambda_{N-1}=~&\omega+\frac{\omega_c^2}{2\omega}\frac{\left(g_{N-1}\omega\right)^2}{\omega^2-\omega_c^2},
\end{align}
\end{subequations}
Below, the 3-mode solution will be applied to the well and barrier regions separately.

\subsection{Two-mode solution at the reactant well}

For the equilibrium reactant well, we can set $\omega_1=\omega$  so the 3-mode potential reads
\begin{align}
\frac{1}{2}\omega_1^2 q_1^2+\frac{1}{2}\omega^2 q_{N-1}^2+\frac{\omega_c^2}{2}\left[q_c+g_{N-1} q_{N-1}+g_1 q_1 \right]^2
\end{align}
We then introduce a N-molecule collective mode $q_N$ as
\begin{align}
q_N=\frac{g_{N-1} q_{N-1} +g_1 q_1 }{\sqrt{g_{N-1}^2+g_1^2}}=\frac{g_{N-1} q_{N-1} + g_1 q_1 }{g_N}
\end{align}
where the collective N-mode coupling strength is
\begin{equation}
 g_N^2=g_{N-1}^2+g_1^2
 \label{gN}
\end{equation}
Then, the 3-mode potential is reduced to 2-mode potential
\begin{align}
U_2= \frac{1}{2}\omega^2 q_N^2+\frac{1}{2}\omega_c^2\left(q_c+g_N q_N \right)^2
\end{align}
which is adopted in the main text.  The 2-mode model is equivalent to the single molecule solution in Eq.~(\ref{A2}) with the replacement
of $g \rightarrow g_N$. As a result, the total frequency at equilibrium is given as
\begin{align}
(\lambda_+ +\lambda_-)_{well} \simeq (\omega+\omega_c)+\frac{\omega_c^2g_N^2}{2(\omega+\omega_c)}
\end{align}
where the last term is the ZPE shift in the reactant well.

\subsection{Three-mode solution at the reactive barrier}

At the reaction barrier (i.e. TS), we identify   $\omega_1^2=-\omega_\ddagger^2$ and $g_1 = g_\dd$ in Eq.~(\ref{A3}). Then, the perturbative expressions
of eigen frequencies become
\begin{subequations}
\begin{align}
\lambda_\dd  \equiv & ~|\lambda_1| =~ \omega_\dd - \frac{\omega_c^2}{2\omega_\dd}\frac{\left(g_\dd \omega_\dd \right)^2}{\omega_\dd^2+\omega_c^2}  \\
\lambda_{b+}  \equiv &~ \lambda_c=~ \omega_c+\frac{\omega_c^3}{2}\left[\frac{g_\ddagger^2}{\omega_c^2+\omega_\ddagger^2}+\frac{g_{N-1}^2}{\omega_c^2-\omega^2}\right] \\
\lambda_{b-}  \equiv &~ \lambda_{N-1} =~\omega+\frac{\omega_c^2}{2\omega}\frac{\left(g_{N-1}\omega\right)^2}{\omega^2-\omega_c^2}
\end{align}
\end{subequations}
where we identify the three frequencies with an unstable mode ($\lambda_\dd$) and a pair of stable modes ($\lambda_{b\pm}$), respectively.

The last two terms of the frequencies of the two stable modes can be combined to yield
\begin{align}
\frac{\omega_c^2g_{N-1}^2}{2}\left[\frac{\omega_c-\omega}{\omega_c^2-\omega^2}\right]=\frac{\omega_c^2g_{N-1}^2}{2}\frac{1}{\omega_c+\omega}
\end{align}
As a result, the frequency shift becomes
\begin{align}
S(N) &= (\lambda_+ +\lambda_-)_{well} - (\lambda_{b+}+\lambda_{b-})_{barrier} \notag \\
& \approx \frac{\omega_c^2}{2}\frac{g^2}{\omega_c+\omega}-\frac{\omega_c^3}{2}\frac{g_\ddagger^2}{\omega_c^2+\omega_\ddagger^2}  \notag \\
& = S(1)
\end{align}
where relation Eq.~(\ref{gN}) is used.
Thus, to leading order of the VSC strength, both the energy shift $S(N)$ and the unstable frequency $\lambda_{\dd}(N)$
 are independent of molecular density in a cavity,
suggesting that cavity-induced correction $\kappa$ is also independent of molecular density.
This conclusion disagrees with experimental measurements, thus implying the failure of the incoherent version of TST  and suggesting the role of coherent cooperativity.   This is a key result of this paper and is further elaborated in  the main text.

To examine the N-dependence within the 3-mode model, we can take a further look at the negative eigenvalue associated with the unstable mode,
\begin{align}
\lambda_\ddagger^2(N) \approx \omega_\ddagger^2+\underbrace{\frac{\left(g_\ddagger g_{N-1}\omega_c^2\right)^2}
{\omega_\ddagger^2+\omega^2}}_{\left(N-1\right)~mode} -
\underbrace{\frac{\left(g_\ddagger\omega_\ddagger\omega_c\right)^2}{\omega_\ddagger^2+\omega_c^2}}_{reative~mode} \approx \lambda_\dd^2(1)
\end{align}
where the second term represents the perturbation due to the collective mode and the third term represents the direct VSC of the reactive mode.
In leading order of the coupling strength, we ignore the second term associated the (N-1) mode and obtain the N-independent expression. Yet, as N increases,
the second term will increase in proportional, thus reducing the suppressing effect of  $\kappa_{GH}$.  This is consistent with
the exact numerical calculation.

\begin{figure}
\centerline{\scalebox{0.4}{\includegraphics[trim=0 0 0 0,clip]{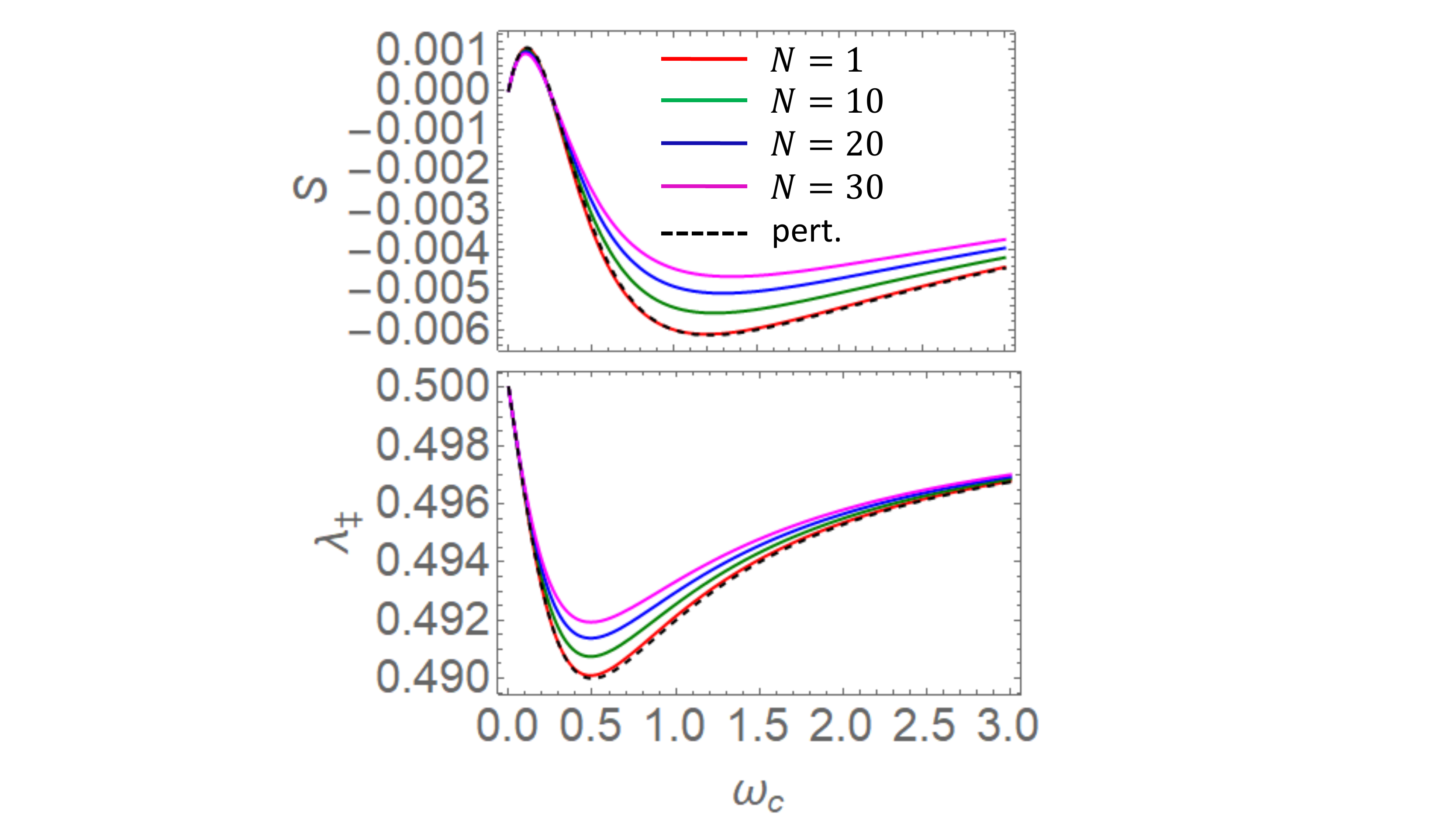}}}
\caption{The ZPE frequency shift $S$ (upper panel) and the unstable eigen frequency $\lambda_\ddagger$ (lower panel)
as a function of $\omega_c$ for different $N$ with
fixed parameters $\eta=\eta_\ddagger=0.1$ and $\omega_{\ddagger}=0.5$.   The N-dependence is introduced in the framework of incoherent TST.
The perturbative results are shown as the dashed line. All physical quantities  are in unit of $\omega$.}
\label{Fig4}
\end{figure}

As a numerical example, in Fig.~S4, we plot the ZPE frequency shift $S$ and the unstable eigen frequency $\lambda_\ddagger$
and compare with the perturbation results (dashed lines). As expected, in the framework of incoherent TST, both quantities
show weak N-dependence and agree well with the perturbation prediction, which is N independent. Again, the classical limit
(i.e. high temperature limit) associated with $\lambda_\dd$ shows the minimal at the barrier resonance whereas
 the quantum limit (i.e., low temperature limit) associated with $S$ shows the minimal near the vibrational resonance.